\documentclass[reprint,floatfix,amsmath,amssymb,showpacs,showkeys]{revtex4-1}

\usepackage[USenglish]{babel}
\usepackage{ucs}				
\usepackage[utf8x]{inputenc}		
\usepackage{graphicx}			
\usepackage[breaklinks=false]{hyperref}			

\begin{document}

\newcommand*{\posv}{\ensuremath{\vektor{x}}}	
\newcommand*{\pos}{\ensuremath{x}}	
\newcommand*{\disv}{\ensuremath{\vektor{r}}}	
\newcommand*{\dis}{\ensuremath{r}}	
\newcommand*{\disdirection}[1]{\ensuremath{r_{#1}}}	
\newcommand*{\kvdirection}{\ensuremath{\hat{\vektor{k}}}}	
\newcommand*{\kvpicked}{\ensuremath{(k,0,\dots,0)\transpose}}	
\newcommand*{\kvarbitrary}{\ensuremath{k\,\kUnitVektor}}	
\newcommand*{\kvabsvalue}{\ensuremath{\lvert\vektor{k}\rvert}}	
\newcommand*{\pprojection}{\ensuremath{p_{1}}}	
\newcommand*{\pradial}{\ensuremath{p_{\text{r}}}}	
\newcommand*{\PsiVektor}{\ensuremath{\Psi}}
\newcommand*{\Psiprojection}{\ensuremath{\Psi_{1}}}	
\newcommand*{\Psiradial}{\ensuremath{\tilde{\Psi}}}	
\newcommand*{\PsiIsotropic}{\ensuremath{\Psi_{1}}}	
\newcommand*{\dTimesTau}{\ensuremath{d\,\tau}}
\newcommand*{\transpose}{\ensuremath{^\mathrm{T}}}
\newcommand*{\intd}[1]{\ensuremath{\text{d}#1}}
\newcommand*{\intdend}[1]{\ensuremath{\;\intd{#1}}}
\newcommand*{\intdbegin}[1]{\ensuremath{\intd{#1}}\;}
\newcommand*{\imag}{\ensuremath{\text{i}}}
\newcommand*{\SingleParticleDiffusivity}{\ensuremath{D_{t}(\tau)}}
\newcommand*{\meanGeneric}[1]{\ensuremath{\langle#1\rangle}}
\newcommand*{\meanD}{\ensuremath{\meanGeneric{D}}}
\newcommand*{\primedindex}[2]{\ensuremath{#1_{#2}^{\prime{}}}}
\newcommand*{\pPrime}[1]{\primedindex{p}{#1}}
\newcommand*{\DPrime}[1]{\primedindex{D}{#1}}
\newcommand*{\varPrime}[1]{\ensuremath{#1^{\prime{}}}}
\newcommand*{\varPPrime}[1]{\ensuremath{#1^{\prime{}\prime{}}}}
\newcommand*{\kTilde}{\ensuremath{\tilde{k}}}
\newcommand*{\pTilde}{\ensuremath{\tilde{p}(D,\kTilde(\tau))}}
\newcommand*{\kmax}{\ensuremath{k_{\text{max}}}}
\newcommand*{\GselfVektor}{\ensuremath{G_{\text{s}}(\disv,\tau)}}
\newcommand*{\GselfSkalar}{\ensuremath{G_{\text{s}}(\dis,\tau)}}
\newcommand*{\Dik}[1]{\ensuremath{D_{#1} \vektor{k}^2}}
\newcommand*{\bigO}[1]{\ensuremath{O\left(#1\right)}}
\newcommand*{\iisf}{\ensuremath{S(\vektor{k},\tau)}}
\newcommand*{\vektor}[1]{\ensuremath{\mathbf{#1}}}
\newcommand*{\kUnitVektor}{\ensuremath{\hat{\vektor{e}}}}
\newcommand*{\mat}[1]{\ensuremath{\mathbf{#1}}}
\newcommand*{\deltadiffusivitiesdefinition}{\ensuremath{\delta\left(D-\SingleParticleDiffusivity\right)}}
\newcommand*{\Secref}[1]{Sec.~\ref{#1}}
\newcommand*{\Eqref}[1]{Eq.~(\ref{#1})}
\newcommand*{\EqsToref}[2]{Eqs.~(\ref{#1}) to (\ref{#2})}
\newcommand*{\EqsAndref}[2]{Eqs.~(\ref{#1}) and (\ref{#2})}
\newcommand*{\Figref}[1]{Fig.~\ref{#1}}
\newcommand*{\Appref}[1]{\appendixname~\ref{#1}}

\newcommand*{\dctitle}{How to compare diffusion processes assessed by single-particle tracking and pulsed field gradient nuclear magnetic resonance}
\newcommand*{\dckeywords}{heterogeneous diffusion, distribution of diffusivities, single-particle tracking, PFG NMR}
\hypersetup{%
	pdftitle	= {\dctitle}, %
	pdfsubject	= {}, %
	pdfauthor	= {Michael Bauer, Rustem Valiullin, G\"unter Radons, and J\"org K\"arger}, %
	pdfkeywords	= {\dckeywords}, %
}


\title{\dctitle}

\author{Michael Bauer}
\affiliation{Institute of Physics, Chemnitz University of Technology, 09107 Chemnitz, Germany}
\author{Rustem Valiullin}
 \email{valiullin@uni-leipzig.de}
\affiliation{Institute of Experimental Physics I, University of Leipzig, 04103 Leipzig, Germany}
\author{Günter Radons}%
 \email{radons@physik.tu-chemnitz.de}
\affiliation{Institute of Physics, Chemnitz University of Technology, 09107 Chemnitz, Germany}
\author{Jörg Kärger}
\affiliation{Institute of Experimental Physics I, University of Leipzig, 04103 Leipzig, Germany}

\date{\today}

\begin{abstract}
Heterogeneous diffusion processes occur in many different fields such as transport in living cells or diffusion in porous media. A characterization of the transport parameters of such processes can be achieved by ensemble-based methods, such as pulsed field gradient nuclear magnetic resonance (PFG NMR), or by trajectory-based methods obtained from single-particle tracking (SPT) experiments. In this paper, we study the general relationship between both methods and its application to heterogeneous systems. We derive analytical expressions for the distribution of diffusivities from SPT and further relate it to NMR spin-echo diffusion attenuation functions. To exemplify the applicability of this approach, we employ a well-established two-region exchange model, which has widely been used in the context of PFG NMR studies of multiphase systems subjected to interphase molecular exchange processes. This type of systems, which can also describe a layered liquid with layer-dependent self-diffusion coefficients, has also recently gained attention in SPT experiments. We reformulate the results of the two-region exchange model in terms of SPT-observables and compare its predictions to that obtained using the exact transformation which we derived.
\end{abstract}


\maketitle

\section{\label{sec:Introduction}Introduction}
Diffusion is one of the omnipresent phenomena in nature involved in most physico-chemical and biological processes \cite{heitjans2005}. Often media, where the molecules perform their chaotic Brownian motion, do include different types of compartments, regions of different densities or domains surrounded by semi-permeable membranes. Diffusion properties in these spatially separated regions may, in general, be different. Altogether, this typically gives rise to very complex processes of diffusive mass transport including regimes of anomalous diffusion. To model such inhomogeneous systems, they may be represented to consist of a number of domains with different local diffusivities subjected to exchange processes between them. The most simple two-phase exchange model with an exponential exchange kernel has often been used to describe experimental results obtained using pulsed field gradient nuclear magnetic resonance (PFG NMR) technique \cite{price2009}. Such examples include, e.g.\ diffusive exchange between two pools of guest molecules in zeolite crystals and surrounding gas atmosphere \cite{krutyeva2008} and between extra- and intracellular water \cite{nilsson2010} in biosystems.

Recently, a new type of experimental approach, namely single-particle tracking (SPT) has emerged \cite{Saxton1997}. It provides an alternative method for studying diffusion processes and for measuring their properties as well as some properties of the surrounding medium \cite{Zurner2007}. In contrast to PFG NMR, where an ensemble of diffusing particles is investigated, SPT only observes individual tracer particles. In particular, fluorescent dye molecules, like rhodamine B, in a solvent, e.g.\ tetrakis(2-ethylhexoxy)-silane (TEHOS), which arranges in ultra-thin liquid layers \cite{yu1999}, are excited by laser radiation. The emitted light of the dyes is captured with a wide-field microscope and recorded by a CCD camera. Hence, the obtained movies show diffusing spots representing a two-dimensional projection of the three-dimensional motion of the dyes. From a statistical point of view, such processes are known as observed diffusion \cite{dembo1986,campillo1989,das2009} or hidden Markov models \cite{rabiner1989,elliott1995} leading in general to the loss of the Markov property. A tracking algorithm detects the positions of the spots and connects them to trajectories \cite{heidernaetsch2009}. A basic quantity for the characterization of diffusion processes is obtained by taking two positions $\posv(t)$ and $\posv(t+\tau)$ from a trajectory separated by a time lag $\tau$ and by considering the rescaled squared displacement $[\posv(t+\tau)-\posv(t)]^{2}/\tau$. This quantity is a local or microscopic diffusivity which fluctuates along a given trajectory or in an ensemble of diffusing particles. It is natural to extract the corresponding distribution of diffusivities from experiments by forming histograms of the observed rescaled squared displacements \cite{bauer2009}. Note that other definitions of diffusivity distributions may be found in the literature \cite{Saxton1997b}. For homogeneous diffusion processes the distribution of diffusivities is independent of the time lag $\tau$, whereas for heterogeneous systems a non-trivial $\tau$-dependence is observed. Therefore in analyzing heterogeneous systems the distribution of diffusivities provides advantages over an analysis via mean squared displacements (msd) because in addition to its mean value it contains all information about the fluctuations \cite{trenkmann2009}. Furthermore, quantities such as the mean diffusion coefficient, obtained as the first moment of the distribution of diffusivities, are well defined, and thus time-dependent diffusion coefficients and their fluctuations can be calculated.

The objective of this work is to investigate the connection between the two different techniques of measuring diffusion. SPT and PFG NMR are clearly related to each other, since both measure displacements of diffusing particles. For instance, the time lag between the observation of two positions in SPT corresponds to the time interval between two gradient pulses in PFG NMR. In both SPT and PFG NMR this time lag $\tau$ is a parameter, which can be tuned by varying the time between snapshots and by altering the temporal distance between gradient pulses, respectively. Furthermore, the signal attenuation in PFG NMR is related to the propagator in Fourier space, from which the distribution of diffusivities can be calculated. At first, it seems to be sufficient to compare the propagators obtained from both types of experiments directly. However, if diffusion processes with heterogeneities or anomalous behavior are considered, access to the propagator will be complicated or even hindered. In such cases, the distribution of diffusivities offers a well-defined analysis of the processes and a comparison of data from the two approaches is feasible. Moreover, it becomes possible to contrast results from time-averaged and ensemble-averaged quantities and detect anomalous diffusion leading to ergodicity breaking as reported recently \cite{lubelski2008}. More generally, an improvement in the analysis of heterogeneous diffusion could benefit from the link between single-particle analysis and ensemble methods. Hence, analytical expressions for one- up to three-dimensional processes are derived which transform PFG NMR signal attenuation into the distribution of single-particle diffusivities from SPT.

For simple systems with heterogeneous diffusion the two-region exchange model of PFG NMR offers an analytical expression for the spin-echo diffusion attenuation \cite{kaerger1988}. In conjunction with our transformation, this model provides an example, where the distribution of single-particle diffusivities can be calculated exactly and also the non-trivial time-lag dependence can be investigated. In this context we consider a two-layer liquid film on a homogeneous surface characterized by two distinct diffusion coefficients \cite{trenkmann2009}. This two-layer system corresponds exactly to the two-region exchange model of PFG NMR. In particular, its condition of exponential waiting times is fulfilled since a change in the diffusion coefficient is possible at any time and independent of the particles' current positions. For a system comprising an arbitrary number of layers, exact asymptotic results for the dispersion of particles in the long-time limit have already been provided \cite{vandenbroeck1984}. We substantiate our findings by analyzing data from simulated single-particle trajectories of heterogeneous diffusion. To evaluate experimental limitations, we study the influence of a signal attenuation bounded to a certain range of $k$-values. The impact on the distribution of single-particle diffusivities will also be pointed out.

The remainder of the paper is organized as follows. In \Secref{sec:Preliminaries} we recall the basic principles of PFG NMR and underline the differences to SPT experiments. In particular, we discuss properties of both approaches and the connection between them. In this context, we introduce the distribution of single-particle diffusivities and provide expressions for the well-known case of homogeneous diffusion. To apply our new concepts to some more elaborated systems, we consider in \Secref{sec:Two-region system} heterogeneous diffusion in two-region systems, where analytical expressions of the PFG NMR signal attenuation exist. We outline the principles of the simulation of such systems in \Secref{sec:Simulation of two-region systems}. In order to provide a simple relation between signal attenuation and distribution of diffusivities, we suggest an approximation in \Secref{sec:Approximation of diffusivity distributions} to avoid the inconvenient Fourier transformation. This approximation is then compared to the exact expressions of the relation in \Secref{sec:Exact relation between signal attenuation and distribution of diffusivities} for simulated data of the two-region system. Finally, in \Secref{sec:Influence of experimentally bounded k} we address the issue of finite intensity of the magnetic field gradient pulses in the PFG NMR experiment and illustrate its influence on our exact transformation into the distribution of diffusivities.

\section{\label{sec:Preliminaries}Signal attenuation and distribution of diffusivities}

Diffusion measurement by PFG NMR is based on observing the transverse magnetization of nuclear spins in a constant magnetic field. Offering the highest sensitivity and occurring in numerous chemical compounds, in most cases the nuclei under study are protons. By superimposing, over two short time intervals, an additional magnetic field with a large gradient, the displacement of the nuclei (and hence of the molecules in which they are contained) in the time span between these two ``gradient pulses'' is recorded in a phase shift of their orientation in the plane perpendicular to the magnetic field with respect to the mean orientation. Hence, the distribution of the diffusion path lengths appears in the distribution of these phase shifts and, consequently, in the vector sum of the magnetic moments of the individual spins, i.e., in the magnetization \cite{price2009,bluemich2005,kimmich2008,valiullin2008}. Since it is this magnetization which is recorded as the NMR signal, molecular diffusion leads to an attenuation of the signal intensity during the PFG NMR experiments which is the larger the larger the displacements in the time interval between these two gradient pulses are.

The signal attenuation from PFG NMR may be shown to obey the relation \cite{price2009,bluemich2005,kaerger1983,kaerger1988}
\begin{equation}
 \label{eqn:signal attenuation Fourier transform of mean propagator}
 \PsiVektor(\tau,\vektor{k})=\int \intdbegin{\disv} p(\disv,\tau) \exp(\imag \vektor{k}\disv)
\end{equation}
with the ensemble-averaged conditional probability density
\begin{equation}
 \label{eqn:mean propagator from ensemble average}
 p(\disv,\tau) = \int\intdbegin{\posv} p(\posv+\disv,\tau \vert \posv)\, p_{0}(\posv)
\text{,}
\end{equation}
where $p(\posv+\disv,\tau \vert \posv)$ is the stationary probability density of a displacement $\disv=(r_1,\dots,r_d)\transpose$ in $d$ dimensions in the time interval $\tau$ and $p_{0}(\posv)$ refers to the equilibrium distribution given by the Boltzmann distribution. Further, $\tau$ is the time interval between the two gradient pulses and represents the diffusion time in the PFG NMR experiment. According to the PFG NMR experiment signal attenuation is measured in the direction of the applied field gradients. Thus, $\vektor{k}$ in \Eqref{eqn:signal attenuation Fourier transform of mean propagator} is given by $\vektor{k} = \kvarbitrary$, where $\kUnitVektor$ denotes the unit vector in that direction. The quantity $k$ is a measure of the intensity of the field gradient pulses. Assuming an isotropic system, without loss of generality, an arbitrary direction $\kvdirection = \kvpicked$ may be considered. Obviously, the scalar product in the exponential of \Eqref{eqn:signal attenuation Fourier transform of mean propagator} picks out only the component $\disdirection{1}$ of the displacement $\disv$. Then, the signal attenuation
\begin{eqnarray}
 \Psiprojection(\tau,k) & = & \PsiVektor(\tau,\kvdirection = \kvpicked) \nonumber \\
 & = & \int\limits_{-\infty}^{+\infty} \intdbegin{\disdirection{1}} \pprojection(\disdirection{1},\tau) \exp(\imag k \disdirection{1})
 \label{eqn:signal attenuation Fourier transform of mean propagator isotropic}
\end{eqnarray}
depends only on scalar values corresponding to the chosen direction and $\pprojection(\disdirection{1},\tau)$ is the projection of the probability density \Eqref{eqn:mean propagator from ensemble average} on the considered direction, given by
\begin{equation}
 \label{eqn:projection of probability density}
\pprojection(\disdirection{1},\tau) = \idotsint \intd{r_2}\cdots\intdbegin{r_d} p(\disv,\tau)
\text{.}
\end{equation}
In NMR $\pprojection(\disdirection{1},\tau)$ in \Eqref{eqn:signal attenuation Fourier transform of mean propagator isotropic} is known as the mean propagator, i.e., the probability density that, during $\tau$, an arbitrarily selected molecule has been shifted over a distance $\disdirection{1}$ in the direction of the applied field gradients. However, it should be noted that for heterogeneous systems, such as systems with regions of different mobility, $\pprojection(\disdirection{1},\tau)$ may not be called propagator since it cannot evolve the system due to the loss of Markovianity. The reason is that, in general, $\pprojection(\disdirection{1},\tau)$ of such systems does not satisfy the Chapman–Kolmogorov equation \cite{gardiner2004}. Non-Markovian behavior, besides others, may also arise in systems which can be modeled by fractional Brownian motion \cite{mandelbrot1968} or by certain fractional diffusion equations \cite{klages2008}. Further, the mean propagator in Fourier space as given by \Eqref{eqn:signal attenuation Fourier transform of mean propagator} corresponds to the incoherent intermediate scattering function. The details of this connection are given in \Appref{sec:Appendix Correspondence iisf and signal attenuation} for clarification.

In contrast to the PFG NMR technique, which is ensemble-based as described above, SPT experiments allow to follow the trace of individual diffusing molecules. Therefore by considering the displacement of a particular molecule in $d$ dimensions it is natural to define a \emph{microscopic} single-particle diffusivity $\SingleParticleDiffusivity$ by the relation
\begin{equation}
 \label{eqn:single particle diffusivity}
 \SingleParticleDiffusivity = [\posv(t+\tau)-\posv(t)]^{2}/(2\dTimesTau)
\text{,}
\end{equation}
where $\posv(t)$ denotes the trajectory of an arbitrary stochastic process. Note that the term ``microscopic'' has been used before by Kusumi and co-workers \cite{kusumi1993} to characterize the short-time behavior of averaged squared displacements equivalent to the small $\tau$ limit of our mean diffusivity defined in \Eqref{eqn:mean diffusivity from first moment} below. In the context of Markovian diffusion processes this limit also corresponds via jump moments to the diffusion terms appearing in Fokker-Planck equations \cite{gardiner2004}. Here we use the term ``microscopic'' in analogy to the statistical physics concept of microstates to distinguish it from ensemble based averages. For a given time lag $\tau$, the microscopic single-particle diffusivity is a fluctuating quantity along a trajectory $\posv(t)$ and we now ask for the probability $p(D)\intd{D}$ that, under the so far considered conditions of normal diffusion, $\SingleParticleDiffusivity$ attains a value in the interval $D\dots D+\intd{D}$. Therefore, the distribution of single-particle diffusivities is defined as
\begin{equation}
\label{eqn:definition of distribution of diffusivities}
p(D,\tau) = \left\langle \deltadiffusivitiesdefinition \right\rangle
\text{,}
\end{equation}
where $\langle \ldots \rangle$ denotes an average, which can be evaluated either as a time-average $\langle \ldots \rangle = \lim_{T \to \infty} 1/T \int_{0}^{T} \dots \intd{t}$, which is accessible by SPT, or, as an ensemble average, appropriate for NMR measurements. Note, however, that in SPT, $T$ is usually limited by the finiteness of the trajectory and complications arise due to the blinking and bleaching of the fluorescent dyes \cite{schuster2007}. However, advanced tracking algorithms in SPT reduce these effects \cite{sbalzarini2005,heidernaetsch2009} and arbitrary time lags $\tau$ between snapshots, which are only limited below by the inverse frame rate of the video microscope, can be accomplished. For experimental SPT data, the distribution of diffusivities is obtained by binning diffusivities into a normalized histogram according to \Eqref{eqn:definition of distribution of diffusivities}.

For ergodic systems, as considered here, time average and ensemble average coincide. By additionally assuming time invariance, \Eqref{eqn:definition of distribution of diffusivities} can be rewritten as
\begin{equation}
\label{eqn:definition of distribution of diffusivities with ensemble-averaged probability density}
p(D,\tau) = \int \intdbegin{\disv} \delta\left(D-\frac{\disv^{2}}{2\dTimesTau}\right) p(\disv,\tau)
\end{equation}
with the probability density \Eqref{eqn:mean propagator from ensemble average} given by $p(\disv,\tau) = \left\langle\delta(\disv-\disv(\tau))\right\rangle$. By performing the angular integration, \Eqref{eqn:definition of distribution of diffusivities} or \eqref{eqn:definition of distribution of diffusivities with ensemble-averaged probability density} can also be expressed as
\begin{equation}
\label{eqn:probability density of single-particle diffusivities}
p(D,\tau) = \int\limits_{0}^{\infty} \intdbegin{\dis} \delta\left(D-\frac{\dis^{2}}{2\dTimesTau}\right) \pradial(\dis,\tau)
\end{equation}
in terms of the radial propagator $\pradial(\dis,\tau)$, which is the probability density $p(\disv,\tau)$ integrated over the surface of a $d$-dimensional sphere with radius $\dis$.

The delta functions in \EqsAndref{eqn:definition of distribution of diffusivities with ensemble-averaged probability density}{eqn:probability density of single-particle diffusivities} simply describe a transformation of the coordinates from displacements to diffusivities. Hence, the distribution of diffusivities is a rescaled version of the propagator. This becomes obvious by expanding for $\dis>0$ the delta function in \Eqref{eqn:probability density of single-particle diffusivities} as $\delta[D-\dis^{2}/(2\dTimesTau)] = \sqrt{\dTimesTau/(2D)} \ \delta[\dis - \sqrt{2\dTimesTau\,D}]$ which yields the relation
\begin{equation}
 \label{eqn:relation of rescaled propagator and distribution of diffusivities}
p(D,\tau) = \sqrt{\frac{\dTimesTau}{2D}}\ \pradial(\sqrt{2\dTimesTau D},\tau)
\text{.}
\end{equation}
Furthermore, it should be noted that the distribution of single-particle diffusivities is closely related to the self part of the van Hove function given in \Appref{sec:Appendix Correspondence iisf and signal attenuation}, which coincides with $p(\disv,\tau)$ given by \Eqref{eqn:mean propagator from ensemble average} for identical particles. Hence, the distribution of diffusivities is also a rescaled version of the van Hove self-correlation function and offers some beneficial properties for our investigations.

The diffusivity $\meanD$ results as the mean of the microscopic single-particle diffusivities \Eqref{eqn:single particle diffusivity}. Therefore, for clarity, we denote it in the following as mean diffusivity. According to the definition of the distribution of diffusivities the mean diffusivity has to obey the relation
\begin{equation}
 \label{eqn:mean diffusivity from first moment}
 \meanGeneric{D(\tau)} = \int\limits_{0}^{\infty} \intdbegin{D} D\,p(D,\tau)
\text{.}
\end{equation}
It is thus obtained as the first moment of the probability density of diffusivities by a well-defined integration, avoiding any numerical fit. Obviously it may also depend on the time lag $\tau$.

In the special case of free diffusion of a particle $\posv(t+\tau) - \posv(t) = \int_{t}^{t+\tau} \intdbegin{t^\prime{}} \vektor{\boldsymbol{\xi}}(t^\prime{})$ is a fluctuating quantity taken from one realization of the Gaussian white noise $\boldsymbol{\xi}(t)$ with variance proportional to the diffusion coefficient. With \Eqref{eqn:signal attenuation Fourier transform of mean propagator isotropic}, the mean propagator and the signal attenuation are seen to be interrelated by Fourier transformation \cite{kaerger1983,kaerger1988}. In the case of normal diffusion in one dimension one has
\begin{equation}
 \label{eqn:mean propagator}
 p_1(\disdirection{1},\tau) = (4 \pi \meanD \tau)^{-1/2} \exp\left(-\frac{\disdirection{1}^{2}}{4 \meanD \tau}\right)
\end{equation}
where $\meanD$ stands for the diffusivity. To avoid confusion we deviated from the usual way of denoting the diffusivity simply by $D$. This is because we use this notation to refer to microscopic single-particle diffusivities $\SingleParticleDiffusivity$. By inserting \Eqref{eqn:mean propagator} into \Eqref{eqn:signal attenuation Fourier transform of mean propagator isotropic} the signal attenuation in PFG NMR experiments is seen to obey the well-known exponential relation
\begin{equation}
 \label{eqn:signal attenuation 1d}
 \Psiprojection(\tau,k) = \exp(-k^{2} \meanD \tau)
\text{.}
\end{equation}

Let us now consider a molecular random walk in a two-dimensional plane. \Eqref{eqn:mean propagator} describes the probability of a molecular displacement in any arbitrarily chosen direction. For the probability that radial molecular displacements are within the interval $\dis\dots \dis+\intd{\dis}$ one obtains, therefore,
\begin{equation}
 \label{eqn:probability of molecular displacement in interval}
 \pradial(\dis,\tau) \intdend{\dis} = \frac{1}{4\pi \meanD \tau} \exp\left(-\frac{\dis^{2}}{4 \meanD \tau}\right) 2\pi \dis \intdend{\dis}
\text{.}
\end{equation}
The mean squared displacement
\begin{equation}
 \label{eqn:mean squared displacement}
 \frac{1}{4\tau} \meanGeneric{\dis^{2}(\tau)} = \frac{1}{4\tau} \int\limits_{0}^{\infty} \intdbegin{\dis} \pradial(\dis,\tau) \dis^{2} = \meanD
\end{equation}
obeys the well-known Einstein relation for normal diffusion in two dimensions. Inserting the corresponding propagator of homogeneous diffusion in two dimensions \Eqref{eqn:probability of molecular displacement in interval} into \Eqref{eqn:definition of distribution of diffusivities with ensemble-averaged probability density} yields the distribution of single-particle diffusivities
\begin{equation}
 \label{eqn:homogeneous probability density of single-particle diffusivities in 2d}
 p(D) = \meanD^{-1} \exp(-D/\meanD)
\text{.}
\end{equation}
In general, for homogeneous diffusion in $d$ dimensions, the distribution of diffusivities is found to be
\begin{equation}
 \label{eqn:homogeneous probability density of single-particle diffusivities general}
 p_{d}(D) = \mathcal{N}_{d}\, \frac{1}{D} \left(\frac{D}{\meanD}\right)^{d/2} \exp\left(-\frac{d}{2} \frac{D}{\meanD}\right)
\text{,}
\end{equation}
where $\mathcal{N}_{d}$ can be obtained from the normalization condition and is explicitly given by
\begin{equation}
 \label{eqn:homogeneous probability density of single-particle diffusivities normalization constant}
\mathcal{N}_{d} = \begin{cases}
	  1/\sqrt{2\pi} & \text{for } d=1 \\
	  1 & \text{for } d=2 \\
	  3\sqrt{3}/\sqrt{2\pi} & \text{for } d=3
\end{cases}
\text{.}
\end{equation}
Since the system is governed by only one diffusion constant, the dependence on $\tau$ vanishes in \Eqref{eqn:homogeneous probability density of single-particle diffusivities general}. However for heterogeneous diffusion, the distribution of single-particle diffusivities additionally depends on the time lag $\tau$. Then, $p(D,\tau)$ cannot generally be expressed by a simple exponential function as in \Eqref{eqn:homogeneous probability density of single-particle diffusivities general}.

By inserting \Eqref{eqn:homogeneous probability density of single-particle diffusivities in 2d}, the first moment of the distribution of diffusivities \Eqref{eqn:mean diffusivity from first moment}
\begin{equation}
 \label{eqn:mean diffusivity from first moment in 2d}
 \int\limits_{0}^{\infty} \intdbegin{D} D/\meanD\, \exp(-D/\meanD) = \meanD
\end{equation}
is easily seen to be fulfilled for homogeneous systems and equals the mean squared displacement obtained in \Eqref{eqn:mean squared displacement}. Hence, with $p(D,\tau)$, which is a rescaled van Hove self-correlation function, it becomes possible to determine the mean diffusion coefficient of the system by ordinary integration.

\begin{figure}[ht]
	\includegraphics{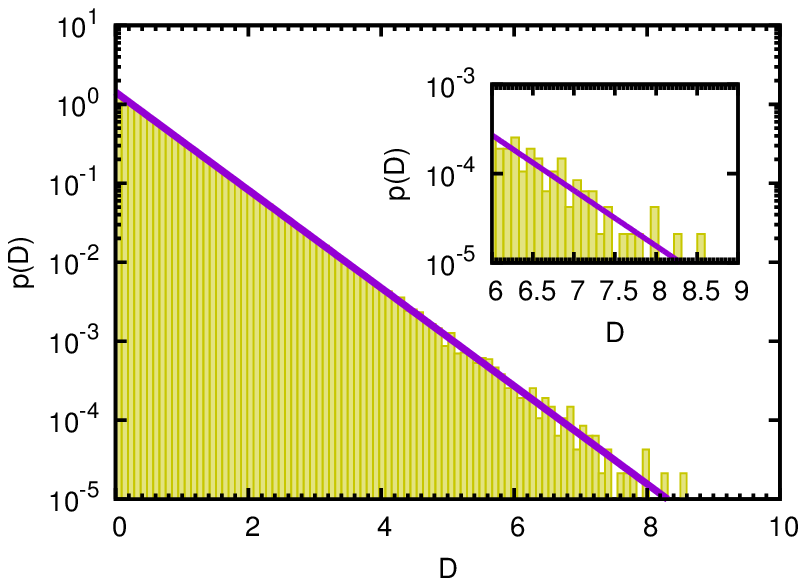}
\caption{\label{fig:pdssd of homogeneous diffusion in two dimensions}Distribution of diffusivities from a simulated trajectory of a homogeneous diffusion process in two dimensions. The distribution agrees well with the exponential behavior expected from \Eqref{eqn:homogeneous probability density of single-particle diffusivities in 2d} and is independent of $\tau$. The inset depicts deviations between simulation and \Eqref{eqn:homogeneous probability density of single-particle diffusivities in 2d} for large $D$ due to insufficient statistics from finite simulation.}
\end{figure}

With \Eqref{eqn:homogeneous probability density of single-particle diffusivities in 2d} for diffusion in two dimensions, the distribution of the single-particle diffusivities in homogeneous systems is seen to result in an exponential. The semi-logarithmic plot of the number of trajectory segments governed by a particular single-particle diffusivity versus these diffusivities is correspondingly expected to yield a straight line. Its negative slope is defined as the reciprocal value of the mean diffusivity. \Figref{fig:pdssd of homogeneous diffusion in two dimensions} depicts the distribution of diffusivities of a homogeneous diffusion process in two dimensions. The data are obtained from simulations of a system with diffusion coefficient $\meanD = 0.7$ and gathered in a normalized histogram. For comparison, the solid line represents the analytical expression \Eqref{eqn:homogeneous probability density of single-particle diffusivities in 2d} and shows a good agreement with the histogram. The inset of \Figref{fig:pdssd of homogeneous diffusion in two dimensions} shows deviations between simulated data and \Eqref{eqn:homogeneous probability density of single-particle diffusivities in 2d} for large $D$ due to insufficient statistics originating from the finite sample in simulation.

It is interesting to note that the shape of the distribution of diffusivities of homogeneous diffusion is similar to that of the attenuation function of PFG NMR diffusion measurements (\Eqref{eqn:signal attenuation 1d}). One has to note, however, that now, in contrast to \Eqref{eqn:signal attenuation 1d}, the mean diffusivity $\meanD$ appears in the denominator of the exponent. From a semi-logarithmic plot of the PFG NMR signal attenuation versus $k^{2}$ the mean diffusivity thus directly results as the slope, rather than its reciprocal value.

In the simple cases of isotropic and homogeneous diffusion both the signal attenuation from PFG NMR and the distribution of diffusivities from SPT resulted in well-known and easily obtainable expressions. In the following we investigate a more elaborated two-region system exhibiting inhomogeneous diffusion.

\section{\label{sec:Two-region system}Heterogeneous diffusion in two-region systems}

Let us now consider molecular diffusion in an isotropic two-region system. With the respective probabilities $\pi_{i}$, the molecules are assumed to propagate with either the diffusivity $D_1$ or $D_2$ and to remain with the mean dwell times $\tau_{m}$ $(m = 1,2)$ in each of these states of mobility. Thus, the observed diffusion process exhibits dynamic heterogeneities emerging as a time-dependent diffusion coefficient due to the exchange of particles between two regions with different diffusion coefficients. For such heterogeneous systems, the behavior of the distribution of single-particle diffusivities, in general, deviates from the mono-exponential decay. This is attributed to a superposition of many different exponentials of type \Eqref{eqn:homogeneous probability density of single-particle diffusivities in 2d} originating from trajectory segments which include layer transitions during the time lag $\tau$. Thus, we denote the distribution of single-particle diffusivities by $p(D,\tau)$ emphasizing its dependence on $\tau$. Further, the superposition and accordingly the characteristics of the distribution of diffusivities strongly depend on the relation of dwell times and the time lag $\tau$ between observed positions \cite{bauer2009}. For short time lags compared to the dwell times the exchange rates are very low. Then, the two diffusion processes can be separated into the two underlying processes. As a result, the probability density is the weighted superposition of the mono-exponential decays belonging to homogeneous diffusion inside each region. In the opposite case, for time lags much larger than both dwell times, the observation only reveals a long-term diffusion process with the mean diffusion coefficient of the system. Hence, the probability density is given by a mono-exponential decay parameterized by this mean diffusivity.

In the case of a two-region system, the PFG NMR spin-echo diffusion attenuation (and hence the Fourier transform of the mean propagator) has been shown to result as a superposition of two terms of the shape of \Eqref{eqn:signal attenuation 1d} \cite{price2009,kaerger1988}:
\begin{eqnarray}
 \PsiIsotropic(\tau,k) & = & \pPrime{1}(k) \exp(-k^{2} \DPrime{1}(k) \tau) \nonumber \\*
	&& + \pPrime{2}(k) \exp(-k^{2} \DPrime{2}(k) \tau)
 \label{eqn:two-region system signal attenuation}
\end{eqnarray}
with
\begin{eqnarray}
 && \DPrime{1,2}(k) = \frac{1}{2} \left(D_1 + D_2 +\frac{1}{k^2}\left(\frac{1}{\tau_1}+\frac{1}{\tau_2}\right)  \phantom{\Biggl(}\right. \nonumber\\*
 && \ \ \mp \left. \left\lbrace\left[D_2 - D_1+\frac{1}{k^2} \left(\frac{1}{\tau_2}-\frac{1}{\tau_1}\right) \right]^2 + \frac{4}{k^4 \tau_1 \tau_2} \right\rbrace^{\frac{1}{2}}\right) \ \ 
 \label{eqn:primed diffusion coefficient}
\end{eqnarray}
\begin{eqnarray}
  \pPrime{1}(k) & = & 1-\pPrime{2}(k) \nonumber\\*
  \pPrime{2}(k) & = & \frac{1}{\DPrime{2}(k)-\DPrime{1}(k)} (\pi_1 D_1 + \pi_2 D_2 -\DPrime{1}(k))
 \label{eqn:primed probability}
\text{.\ \ }
\end{eqnarray}

It should be noted that the primed quantities in \EqsAndref{eqn:primed diffusion coefficient}{eqn:primed probability} depend on the intensity of the magnetic field gradient being related to $k$ and, thus, on the Fourier coordinate. Therefore, \Eqref{eqn:two-region system signal attenuation} cannot be considered as a superposition of separated populations of the two regions. It is rather the total interference of spin-echo attenuations observed from both regions. Further, the initial condition of a process described by \EqsToref{eqn:two-region system signal attenuation}{eqn:primed probability} has to be chosen in such a manner that for the initial time $t = 0$ the particles are located at a given position $\posv$ and are already distributed stationarily between the regions. This is obvious since neither $\pPrime{1}(k)$ nor $\pPrime{2}(k)$ depends on $t$ which would be necessary to converge to the stationary distribution. For any other initial distribution \Eqref{eqn:two-region system signal attenuation} will only be valid in the limit of $t \to \infty$.

The signal attenuation can also be considered for the limiting cases. For $\tau \to 0$, i.e., $\tau \ll \tau_{1},\tau_{2}$, the signal attenuation
\begin{equation}
 \label{eqn:signal attenuation for tau to 0}
\PsiIsotropic(\tau,k) = \pi_{1} \exp(-k^2 D_{1} \tau) + \pi_{2} \exp(-k^2 D_{2} \tau)
\end{equation}
decomposes into the superposition of two signal attenuations corresponding to each region. As discussed, two completely separated diffusion processes are observed. Hence, the inverse Fourier transformation leads to a superposition of the distribution of diffusivities of each region. In contrast, for $\tau \to \infty$, i.e., $\tau \gg \tau_{1},\tau_{2}$, the mixing of the two regions leads to the observation of an effective mean diffusion process with a signal attenuation
\begin{equation}
 \label{eqn:signal attenuation for tau to infinity}
\PsiIsotropic(\tau,k) = \exp(-k^2 (\pi_{1} D_{1} + \pi_{2} D_{2}) \tau)
\end{equation}
containing the mean diffusion coefficient. Analogously, its inverse Fourier transform, i.e., the distribution of diffusivities, is only characterized by the mean diffusion coefficient $\meanD=\pi_{1} D_{1} + \pi_{2} D_{2}$. A detailed deviation of the limiting cases is given in \Appref{sec:Appendix Exact transformation of limiting cases}.

\section{\label{sec:Simulation of two-region systems}Simulation of two-region systems}
In order to simulate heterogeneous diffusion we consider a system with two regions where particles propagate with different diffusivities and can change their state of mobility. Following the experiment with rhodamine in TEHOS \cite{Schob2006,trenkmann2009}, this two-region system is modeled by a bi-layer system with layer-dependent diffusion coefficients $D_1$ and $D_2$, respectively. Such processes can formally be described as composite Markov processes \cite{kampen1992} or equivalently as multistate random walks  \cite{haus1987,weiss1994}, which are known to be widely applicable. A recent biophysical application consists of changes in the diffusive behavior of molecules in membranes due to random changes of the molecules’ conformation \cite{malchus2010}. In the case of two states or regions the probability density of finding the particle at position $\posv$ at time $t$ is determined by the evolution equations
\begin{eqnarray}
\frac{\partial}{\partial t} \hat{p}_{1}(\posv,t) & = & w_{12} \hat{p}_{2}(\posv,t) - w_{21} \hat{p}_{1}(\posv,t) + D_{1} \nabla^{2} \hat{p}_{1}(\posv,t) \nonumber \\*
\frac{\partial}{\partial t} \hat{p}_{2}(\posv,t) & = & w_{21} \hat{p}_{1}(\posv,t) - w_{12} \hat{p}_{2}(\posv,t) + D_{2} \nabla^{2} \hat{p}_{2}(\posv,t) \nonumber \\*
 \label{eqn:equations of motion}
\end{eqnarray}
for each region with corresponding diffusion coefficients $D_1$ and $D_2$. Within each region the motion of the molecules is accomplished by ordinary two-dimensional diffusion, i.e., random walkers experiencing shifts of the positions distributed according to a Gaussian with a variance defined by the diffusion coefficient in the region. The exchange between these two diffusive regions is simulated by a jump process governed by a master equation with jump rates $w_{nm}$, which describe a transition from region $m$ to $n$ $(m,n=1,2)$. The inverse of the jump rates $w_{nm}$ yields the mean dwell time $\tau_{m}$
\begin{equation}
 \label{eqn:dwell time}
 \tau_1 = \frac{1}{w_{21}} \text{ and } \tau_2 = \frac{1}{w_{12}}
\end{equation}
for which particles remain in region $m$. Further, the stationary distribution between the regions
\begin{equation}
 \label{eqn:stationary distribution between regions}
 \pi_1 = \frac{w_{12}}{w_{12}+w_{21}} \text{ and } \pi_2 = \frac{w_{21}}{w_{12}+w_{21}}
\end{equation}
is also dictated by the jump rates. With the stationary distribution the mean diffusion coefficient of the two-region system is given by
\begin{equation}
 \label{eqn:mean diffusion coefficient from stationary distribution}
 \meanD = \pi_{1} D_{1} + \pi_{2} D_{2}
\text{,}
\end{equation}
which is the weighted average of the diffusion coefficients belonging to each region \cite{vandenbroeck1984}.

\begin{figure}[ht]
	\includegraphics{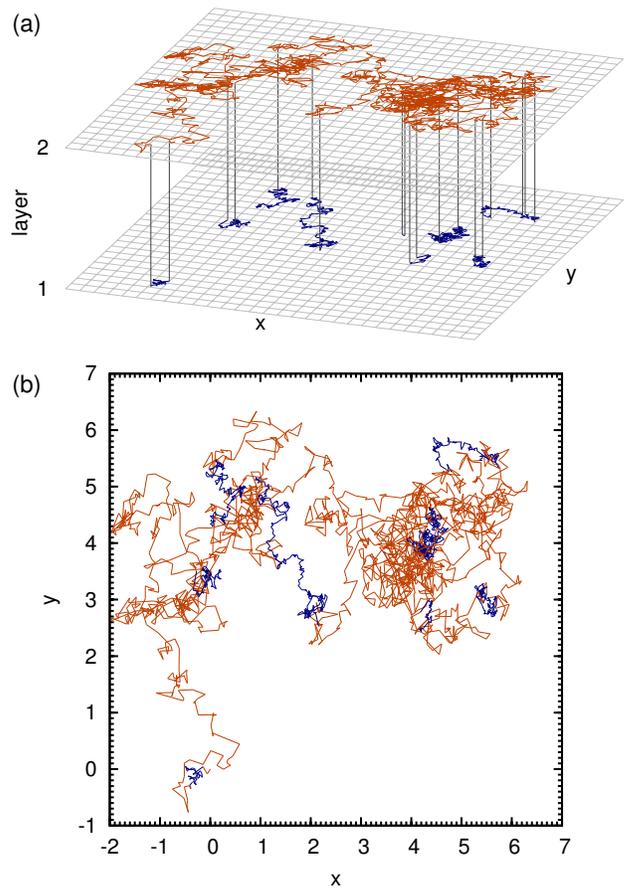}
\caption{\label{fig:trajectory}Single-particle trajectory from simulation of diffusion in a bi-layer system. (a) The particle performs diffusion with corresponding diffusion coefficients and jumps between the layers. (b)
Projection of the trajectory shown in (a) onto the $x$-$y$-plane as usually observed by single-particle tracking. Information of the layer and the corresponding diffusion coefficient is lost in the projection and can only be identified due to the color code.}
\end{figure}

To investigate the effects of heterogeneous diffusion, simulation of the two-region system is performed with the following system parameters. The diffusion coefficients within each of the two regions are given by $D_{1}=0.1$ and $D_{2}=1.0$. The jump rates $w_{21}=8$ and $w_{12}=4$ yield the dwell times $\tau_{1}=0.125$ and $\tau_{2}=0.25$, respectively. Hence, the stationary distribution between the regions results in $\pi_{1}=\frac{1}{3}$ and $\pi_{2}=\frac{2}{3}$ and a mean diffusion coefficient $\meanD=0.7$ is obtained. The length of the time step in the simulation is chosen to be $\Delta t=0.01$, which is much smaller than the dwell times to ensure diffusive motion of the particles within the regions.

Simulation of \Eqref{eqn:equations of motion} is depicted in \Figref{fig:trajectory}~(a). It shows the trajectory of a particle in a bi-layer system, where the particle jumps between the layers. In each layer, diffusion is governed by a different diffusion coefficient denoted by the color of the trajectory segments. Since in experiments with video microscopy only a two-dimensional projection of the process is observed, the trajectory is projected onto the $x$-$y$-plane in \Figref{fig:trajectory}~(b). As a consequence, information about the layer is obscured and can only be identified due to the color coding in the figure. Hence, in the projection it is unknown which diffusion coefficient currently governs the process. A description of such observed diffusion processes by the Fokker-Planck equation with time-dependent diffusion coefficient would become possible if all trajectories jump synchronously. Since in our bi-layer system the particles move independently, the process is more complicated. As a result of the projection, the observed process does not possess the Markov property anymore since, in general, the Chapman–Kolmogorov equation cannot be satisfied. The simulation provides an approach to study properties of an $N$-layer system, which is closely related to a system where the diffusion coefficient varies continuously with the $z$-coordinate.

To avoid transient behavior in our simulation, the particle positions are initialized with their corresponding stationary distributions between the layers given by $\pi_i$. It should be noted, however, that experimental results will be influenced by such transient effects if the tracer molecules require a sufficiently long time to distribute between the layers of the solvent. On the other hand, such slow relaxation is related to low exchange rates leading to almost complete separation of the two diffusive regions \cite{bauer2009}. This would allow for an appropriate bi-exponential fit of our distribution of diffusivities although the weights do not correspond to the stationary distributions yet.

To investigate the connection between spin-echo signal diffusion attenuation, as measured by PFG NMR, and distribution of single-particle diffusivities, as assessed by SPT, we simulated one particle. Next, we recorded squared displacements along the simulated trajectory of $10^{7}$ time steps. The squared displacements are calculated from the changes of the particle positions and are divided by the time lag $\tau$ elapsed between the observations of the two positions. Hence, we obtain scaled squared displacement with the dimension of a diffusion coefficient. The thus obtained diffusivity is a fluctuating quantity along a trajectory. Finally, we gather them in a histogram counting their occurrences. The histogram contains data from a moving-time average since the diffusivities originate from single trajectories. Note that for ergodic systems ensemble averaging will yield identical results. After normalizing the histogram we obtain a probability density referred to as the distribution of diffusivities. The distribution of diffusivities contains all information about the diffusivities of the process and their fluctuations. Following the experiment, only a fraction of the time steps is available for the distribution depending on the selected time lag. Thus, our resulting distributions of diffusivities depicted in log-linear plots have their lower boundary at $10^{-3}$ since data below suffer from insufficient statistics.

\section{\label{sec:Approximation of diffusivity distributions}Approximation of diffusivity distributions}

Since an exact relation of the PFG NMR signal attenuations to distributions of diffusivities requires inverse Fourier transformation we are now going to use the set of \EqsToref{eqn:two-region system signal attenuation}{eqn:primed probability} for an approximation of the probability distribution of the single-particle diffusivities in a two-region system. We proceed in analogy with our treatment of the simple system with only one (mean) diffusivity. In either case the information about the probability distribution $p(D,\tau)$ of the single-particle diffusivities $D$ is clearly contained in the propagator. For the system with one diffusivity this propagator is given by \Eqref{eqn:mean propagator}. Its Fourier transform (\Eqref{eqn:signal attenuation 1d}, which is nothing else than the PFG NMR spin-echo diffusion attenuation curve) was found to coincide with the shape of the probability distribution of the single-particle diffusivities (\Eqref{eqn:homogeneous probability density of single-particle diffusivities in 2d}) with the only difference that the mean diffusivity, which represents the slope in the semi-logarithmic attenuation plots, appears in the denominator of the exponent in the distribution function $p(D)$.

In the two-region system, the PFG NMR spin-echo diffusion attenuation (and hence the Fourier transform of the propagator) is now found to be given by two exponentials (\Eqref{eqn:two-region system signal attenuation}) of the form of \Eqref{eqn:signal attenuation 1d}. Formally we may refer, therefore, to two populations with the relative weights $\pPrime{i}$ and the effective (mean) diffusivities $\DPrime{i}$ as quantified by \EqsAndref{eqn:primed diffusion coefficient}{eqn:primed probability}. Following the analogy of our simple initial system, as a first attempt, the resulting probability function of the single-particle diffusivities may be approximated by a corresponding superposition of two exponentials of the type of \Eqref{eqn:homogeneous probability density of single-particle diffusivities in 2d}
\begin{eqnarray}
 p(D,\tau) \simeq \pTilde & = & \pPrime{1}(\kTilde)\frac{1}{\DPrime{1}(\kTilde)} \exp(-D/\DPrime{1}(\kTilde)) \nonumber \\*
						&& + \pPrime{2}(\kTilde)\frac{1}{\DPrime{2}(\kTilde)} \exp(-D/\DPrime{2}(\kTilde)) \nonumber \\*
  \label{eqn:two-region system naive approximation}
\end{eqnarray}
with the parameters $\pPrime{i}(\kTilde)$ and $\DPrime{i}(\kTilde)$ as given by \EqsAndref{eqn:primed diffusion coefficient}{eqn:primed probability}. Since this approximation avoids Fourier transformation, a proper $\tau$-dependence of $\kTilde$ has to be chosen for the primed quantities. It should be noted that the transformation of \Eqref{eqn:two-region system signal attenuation} from Fourier space will only result in a superposition of two exponentials in real space if the primed quantities in Fourier space are independent of $\kTilde$. Hence, \Eqref{eqn:two-region system naive approximation} could only serve as a rough approximation of the observed process. However, inserting \Eqref{eqn:two-region system naive approximation} into \Eqref{eqn:mean diffusivity from first moment}, the mean diffusivity of the two-region system results in
\begin{equation}
 \label{eqn:two-region system mean diffusivity}
 \meanD = \pPrime{1}(\kTilde) \DPrime{1}(\kTilde) + \pPrime{2}(\kTilde)\DPrime{2}(\kTilde)= \pi_1 D_1 + \pi_2 D_2
\end{equation}
with the second equality resulting from the application of \EqsAndref{eqn:primed diffusion coefficient}{eqn:primed probability}. This is exactly the result which is well-known \cite{vandenbroeck1984} and it should be noted that it does not depend on $\tau$.

Further on, we may consider the limiting cases $\kTilde \to 0$ and $\kTilde \to \infty$ which can be translated to $\dis \to \infty$ and $\dis \to 0$, respectively. Intuitively, large displacements $\dis \to \infty$ are related to long observation times $\tau \to \infty$ and vice versa. This relation is substantiated by keeping $\kTilde^{2} \tau$ constant (see also \Eqref{eqn:k estimation for non-limiting cases}) where $\kTilde \to 0$ corresponds to $\tau \to \infty$ and vice versa. Due to this, the respective limits of $\pTilde$ and $p(D,\tau)$ should coincide. As a result we obtain the expected expressions
\begin{eqnarray}
 && \lim_{\kTilde \to \infty} \pTilde = \lim_{\tau \to 0} p(D,\tau) \nonumber \\*
 \label{eqn:limiting case k to infinity}
 && = \pi_1 D_1^{-1} \exp(-D/D_1) + \pi_2 D_2^{-1} \exp(-D/D_2)
\end{eqnarray}
and
\begin{eqnarray}
 && \lim_{\kTilde \to 0} \pTilde = \lim_{\tau \to \infty} p(D,\tau) \nonumber \\*
 \label{eqn:limiting case k to zero}
 && = \meanD^{-1} \exp(-D/\meanD)
\text{.}
\end{eqnarray}

Since the diffusivities and probabilities $\DPrime{i}$ and $\pPrime{i}$ occurring in \EqsToref{eqn:two-region system signal attenuation}{eqn:primed probability} depend on the Fourier coordinate $\kTilde$, we have referred to the probability density in this context as an approximated one, $\pTilde$. Hence, \Eqref{eqn:two-region system naive approximation} in the given notation is unable to provide an approximation of the probability distribution function of the single-particle diffusivities over the whole diffusivity scale. This is in perfect agreement with previous results \cite{bauer2009} where it has been shown that the distribution of diffusivities, in general, cannot be represented by a weighted superposition of the underlying homogeneous diffusion processes. However, such an approximation of the probability density might become possible by inserting an appropriately selected value for the Fourier coordinate. As a first trial, one may put
\begin{equation}
 \label{eqn:k estimation for non-limiting cases}
 \kTilde^{-2} = \meanD \tau
\text{,}
\end{equation}
which ensures highest sensitivity with respect to the space scale covered during the experiments. Note that in PFG NMR experiments the exponent in the signal attenuation \Eqref{eqn:signal attenuation 1d} is of the order of 1, which yields an easily observable PFG NMR spin-echo diffusion attenuation.

\begin{figure}[ht]
	\includegraphics{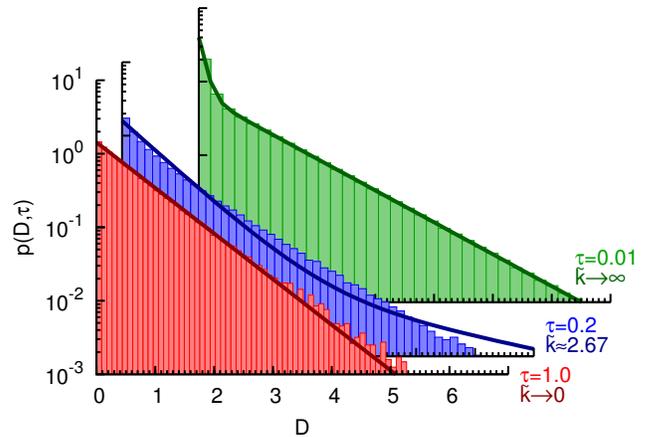}
\caption{\label{fig:pdssd vs estimate for different k}Comparison of distribution of diffusivities (colored histograms) from a simulated two-dimensional trajectory with numerical approximation via \Eqref{eqn:two-region system naive approximation} (solid lines) of a two-region system for time lags $\tau=0.01$, $0.2$ and $1.0$ and mean dwell times of $\tau_1=0.125$ and $\tau_2=0.25$. The limiting cases of $\kTilde \to 0$ and $\kTilde \to \infty$ approximate the simulated data reasonably. However, for $\tau=0.2$ in the order of the dwell times an intermediate $\kTilde\approx2.67$, as suggested in \Eqref{eqn:k estimation for non-limiting cases}, does not approximate the density sufficiently.}
\end{figure}

\Figref{fig:pdssd vs estimate for different k} depicts the distribution of diffusivities from a simulated two-dimensional trajectory in a two-region system with mean dwell times $\tau_1=0.125$ and $\tau_2=0.25$ for three time lags $\tau=0.01$, $0.2$ and $1.0$. Further, the approximation of the distribution of diffusivities from \Eqref{eqn:two-region system naive approximation} is investigated for corresponding $\kTilde$. Thus, the limiting case of completely separated diffusion processes found for $\tau \to 0$ is simulated with $\tau=0.01 \ll \tau_1,\tau_2$ and compared with \Eqref{eqn:two-region system naive approximation} for $\kTilde \to \infty$, i.e., \Eqref{eqn:limiting case k to infinity}. On the other hand, the second limiting case of mean diffusion emerging for $\tau \to \infty$ is obtained from simulation with $\tau=1.0 \gg \tau_1,\tau_2$ and comparison with \Eqref{eqn:two-region system naive approximation} for $\kTilde \to 0$, i.e., \Eqref{eqn:limiting case k to zero}. \Figref{fig:pdssd vs estimate for different k} clearly shows that simulated data from both limiting cases are recovered reasonably by \Eqref{eqn:two-region system naive approximation} for corresponding $\kTilde$. In contrast, the distribution of diffusivities reveals a more complicated behavior in the intermediate exchange regime between the limiting cases. Since the time lag $\tau=0.2$ is in the order of the mean dwell times, neither a mean diffusion process nor a weighted superposition of completely separated processes is observed. In particular, the distribution cannot be approximated by \Eqref{eqn:two-region system naive approximation} with a given $\kTilde(\tau)$. This is obvious, since with such an estimate of $\kTilde$ the dependence on $k$ of the primed quantities \EqsToref{eqn:primed diffusion coefficient}{eqn:primed probability} in Fourier space is neglected. Then, the inverse Fourier transformation of \Eqref{eqn:two-region system signal attenuation} as well as the transformation to the distribution of diffusivities would yield a simple superposition of two exponentials again. In general, this does not provide appropriate results for arbitrary dwell times and time lags \cite{bauer2009}. As a consequence, a general expression requires inverse Fourier transformation of the PFG NMR attenuation curve.

\section{\label{sec:Exact relation between signal attenuation and distribution of diffusivities}Exact relation between signal attenuation and distribution of diffusivities}

In \Secref{sec:Approximation of diffusivity distributions}, the approximation of the distribution of diffusivities by \Eqref{eqn:two-region system naive approximation} was shown to reproduce the limiting cases of time lag $\tau$ as well as the correct mean value. Cases in between the limits did not deliver appropriate results. In order to produce proper results for arbitrary $\tau$ we derive general formulae for the transformation of PFG NMR signal attenuations to distributions of single-particle diffusivities.

Quite formally two steps have to be accomplished to derive a general expression of $p(D,\tau)$ from $\PsiVektor(\tau,\vektor{k})$. As a first step, inverse Fourier transformation of \Eqref{eqn:signal attenuation Fourier transform of mean propagator} yields the propagator in real space. Further, the shift $\disv$ between positions, as given by the propagator, can be translated into diffusivities via scaled squared displacements leading to the distribution of diffusivities as defined in \Eqref{eqn:definition of distribution of diffusivities with ensemble-averaged probability density}.

The two steps can be combined to directly obtain the probability density from signal attenuation. Depending on dimensionality $d$, the distribution of diffusivities is given by
\begin{eqnarray}
p(D,\tau) = \int && \intdbegin{\disv} \delta\left(D-\frac{\disv^{2}}{2d\;\tau}\right) \nonumber \\*
	 && \times \frac{1}{(2\pi)^{d}} \int\intdbegin{\vektor{k}} \PsiVektor(\tau,\vektor{k}) \exp(-\imag\vektor{k}\disv)
 \label{eqn:transformation general}
\text{.}
\end{eqnarray}
With the rescaled coordinates
\begin{equation}
 \varPrime{\disv} = \frac{\disv}{\sqrt{2d\;\tau}} \text{ and } \varPrime{\vektor{k}} = \vektor{k} \sqrt{2d\;\tau}
\end{equation}
it is further simplified to
\begin{equation}
\label{eqn:transformation general simplified}
 p(D,\tau) = \int\intdbegin{\varPrime{\vektor{k}}} \PsiVektor\left(\tau,\frac{\varPrime{\vektor{k}}}{\sqrt{2d\;\tau}}\right) \frac{1}{(2\pi)^{d}} S_{d}(\varPrime{\vektor{k}},D)
\text{,}
\end{equation}
with $S_{d}(\vektor{k},D)$ being the Fourier transform of a uniform density on the surface of a $d$-dimensional sphere of radius $\sqrt{D}$
\begin{equation}
\label{eqn:delta function on hypersphere}
S_{d}(\vektor{k},D) = \int\intdbegin{\disv} \delta(D-\disv^{2}) \exp(-\imag\vektor{k}\disv)
\text{.}
\end{equation}
Since \Eqref{eqn:delta function on hypersphere} can be expressed analytically \cite{vembu1961} by
\begin{equation}
 \label{eqn:transformation generic}
 S_{d}(\vektor{k},D) = \pi^{a+1} D^a 2^a J_{a}(\kvabsvalue \sqrt{D}) (\kvabsvalue \sqrt{D})^{-a}
\end{equation}
with $a=d/2-1$ and $J_{a}(x)$ denoting the Bessel function of the first kind, the exact transformation of signal attenuations $\PsiVektor(\tau,\vektor{k})$ to distributions of diffusivities $p(D,\tau)$ is accomplished without applying an inverse Fourier transformation.

For isotropic systems, the signal attenuation $\PsiVektor(\tau,\vektor{k})$ depends only on the absolute value of $\vektor{k}$, i.e., the radial intensity of the field gradient $k$. Without loss of generality, an arbitrary direction $\vektor{k}=\kvpicked$ may be considered and the corresponding signal attenuation is denoted by $\PsiIsotropic(\tau,k)=\PsiVektor(\tau,\vektor{k}=\kvpicked)$. Then the following expressions are obtained for the distribution of diffusivities depending on the dimensionality of the system. For one-dimensional systems \Eqref{eqn:transformation general} reduces to
\begin{equation}
 \label{eqn:transformation 1d}
 p(D,\tau) = \frac{1}{\pi\sqrt{D}} \int\limits_{0}^{\infty} \intdbegin{k} \PsiIsotropic\left(\tau,\frac{k}{\sqrt{2\tau}}\right) \cos(k\sqrt{D})
\text{.}
\end{equation}
The transformation for $d=2$ can be written as
\begin{equation}
 \label{eqn:transformation 2d}
 p(D,\tau) = \frac{1}{2} \int\limits_{0}^{\infty} \intdbegin{k} \PsiIsotropic\left(\tau,\frac{k}{\sqrt{4\tau}}\right) k J_{0}(k\sqrt{D})
\text{,}
\end{equation}
and for $d=3$ one obtains
\begin{equation}
 \label{eqn:transformation 3d}
 p(D,\tau) = \frac{1}{\pi} \int\limits_{0}^{\infty} \intdbegin{k} \PsiIsotropic\left(\tau,\frac{k}{\sqrt{6\tau}}\right) k \sin(k\sqrt{D})
\end{equation}
using polar and spherical coordinates, respectively. The given transformations move the whole dependence on time lag $\tau$ to the signal attenuation. This is achieved by rescaling the $k$ coordinate by $\sqrt{2d\;\tau}$.

Hence, a signal attenuation of an ensemble diffusing in a two-dimensional plane measured by PFG NMR is transformed into a distribution of single-particles diffusivities via \Eqref{eqn:transformation 2d}. For homogeneous diffusion \Eqref{eqn:transformation 2d} yields the expected probability of single-particle diffusivities \Eqref{eqn:homogeneous probability density of single-particle diffusivities in 2d} by inserting the simple exponential relation \Eqref{eqn:signal attenuation 1d} as signal attenuation.

Furthermore, the limiting cases of time lag $\tau$ are reproduced exactly by the presented transformations \EqsToref{eqn:transformation 1d}{eqn:transformation 3d}: For $\tau \to 0$ the distribution of single-particle diffusivities for the given dimensionality results in the superposition of two respective distributions \Eqref{eqn:homogeneous probability density of single-particle diffusivities general} denoting two separated, homogeneous diffusion processes. On the other hand, for $\tau \to \infty$, the resulting distribution of single-particle diffusivities for the given dimensionality is also of type \Eqref{eqn:homogeneous probability density of single-particle diffusivities general}, respectively, and depends only on the mean diffusion coefficient of the system. A detailed derivation of the limiting cases is given in \Appref{sec:Appendix Exact transformation of limiting cases}.

\begin{figure}[ht]
	\includegraphics{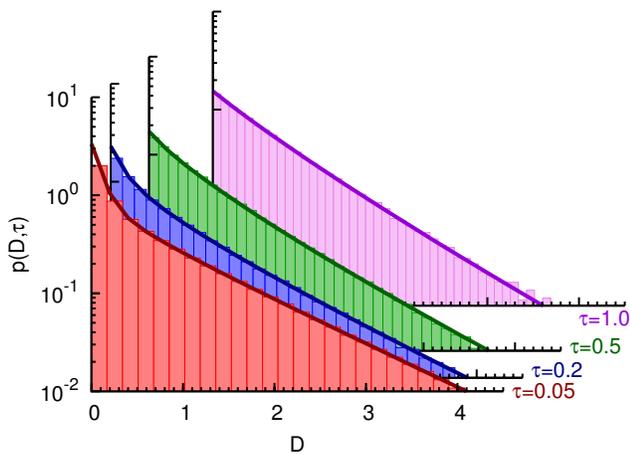}
\caption{\label{fig:pdssd vs complete inverse Fourier}Comparison of distributions of single-particle diffusivities from a simulated two-dimensional trajectory (colored histograms) with distributions obtained by applying \Eqref{eqn:transformation 2d} for an exact transformation of the PFG NMR spin-echo signal diffusion attenuation \Eqref{eqn:two-region system signal attenuation} of a two-region system for time lags $\tau=0.05$, $0.2$, $0.5$ and $1.0$ and mean dwell times $\tau_1=0.125$ and $\tau_2=0.25$ (solid lines). The data agree well with each other for each $\tau$. Further, the dependence on $\tau$ is apparent, which is typical for diffusion in heterogeneous media.}
\end{figure}

To examine the transformations, the same parameters, for which the approximation via \Eqref{eqn:two-region system naive approximation} failed, are used again, now applying \Eqref{eqn:transformation 2d} for an exact transformation of the PFG NMR signal attenuation relation, \Eqref{eqn:two-region system signal attenuation}, into the distribution of single-particle diffusivities. The results are depicted in \Figref{fig:pdssd vs complete inverse Fourier} and again the distribution of diffusivities from a simulated two-dimensional trajectory is given for comparison. For each of the chosen $\tau=0.05$, $0.2$, $0.5$ and $1.0$ a perfect agreement is obvious, confirming the relation between the two approaches. Moreover, \Figref{fig:pdssd vs complete inverse Fourier} clearly illustrates how the distribution of diffusivities depends on $\tau$ and reveals a transition from a non-exponential behavior to a mono-exponential decay. For small $\tau$ corresponding to diffusion in separated regions it deviates considerably from a mono-exponential behavior. However, for long-term observations ($\tau \to \infty$) only a mean diffusion process is observed due to averaging of the motion in both regions. Consequently, this yields a mono-exponential decay of the distribution of diffusivities. This transition reveals the heterogeneity of the diffusion process \cite{bauer2009}. Hence, in order to characterize diffusive motion the distribution of diffusivities has to be investigated for its dependence on the time lag $\tau$.

\section{\label{sec:Influence of experimentally bounded k}Influence of experimentally bounded \texorpdfstring{$k$}{k}}

PFG NMR spin-echo diffusion attenuation functions can only be measured up to a finite intensity $k$ of the magnetic field gradient pulses. However, to generate the distribution of diffusivities exactly, the signal attenuation has to be given over the whole intensity scale. Hence, the effect of an experimentally bounded Fourier coordinate $k$ has to be considered.

\begin{figure}[ht]
	\includegraphics{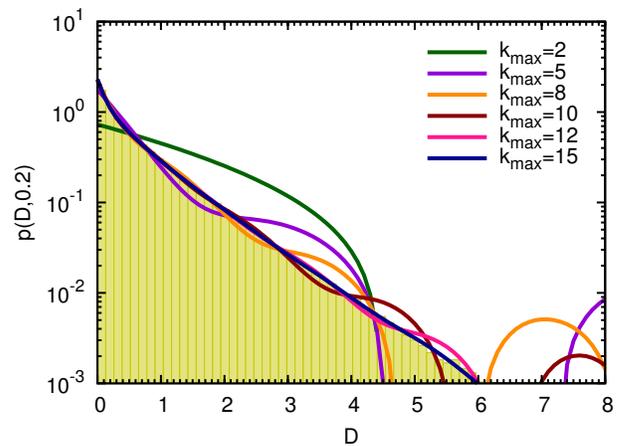}
\caption{\label{fig:pdssd vs inverse Fourier with bounded k}Transformation \Eqref{eqn:transformation 2d} of PFG NMR spin-echo signal diffusion attenuation by integration up to $\kmax$ (solid lines) due to experimentally bounded intensity $k$ of the field gradient pulses. The distribution of single-particle diffusivities (colored histogram) from a simulated two-dimensional trajectory will only be obtained reasonably if $k$ is given over the whole intensity scale. For smaller intervals of $k$ deviations become clearly visible as well as oscillations introduced by the inverse Fourier transformation.}
\end{figure}

\Figref{fig:pdssd vs inverse Fourier with bounded k} illustrates the influence of finite $k$ on the distribution of single-particle diffusivities obtained for $\tau=0.2$. If with the maximal applied $\kmax$ the respective spin-echo signal is not sufficiently attenuated, the transformation of the signal attenuation from a finite interval will yield significant deviations from the expected probability distribution. As a consequence, the first moment, i.e., the mean diffusion coefficient of the system, is altered accordingly. Furthermore, due to the bounded signal attenuation, the inverse Fourier transformation introduces oscillations since only a limited range of the spectrum contributes to the values in real space. The reason is the integrand in \EqsToref{eqn:transformation 1d}{eqn:transformation 3d} which will only vanish for large $k$ if $\PsiIsotropic$ decays faster than the remainder.

This effect may clearly be identified in \Figref{fig:pdssd vs inverse Fourier with bounded k}. In order to obtain reasonable results, the signal must be attenuated to a sufficient extent. Simulated data of two-dimensional diffusion processes have shown that the attenuation should fall below $10^{-4}$ of its maximum at $\kmax$ to suppress oscillations. This has to be considered when dealing with experimental data.

The necessity of fast decaying $\PsiIsotropic$ becomes especially important for large time lags $\tau$. In the case of small time lags $\tau \to 0$ our rescaling of the $k$ coordinate in \EqsToref{eqn:transformation 1d}{eqn:transformation 3d} leads to $k/\sqrt{\tau} \to \infty$ in the second argument of $\PsiIsotropic(\tau,k/\sqrt{2d\;\tau})$. Thus, for small $\tau$, signal attenuation becomes more pronounced and reduces the influence of the bounded $k$. Moreover, signal attenuation is closely related to the incoherent structure factor \cite{fleischer1994}, as demonstrated in \Appref{sec:Appendix Correspondence iisf and signal attenuation}, dealing with similar limitations. A possible solution is to split the integral into two parts, integrating numerically up to the experimental limit $\kmax$ and assuming an analytical expression for the remaining part.

\begin{figure}[ht]
	\includegraphics{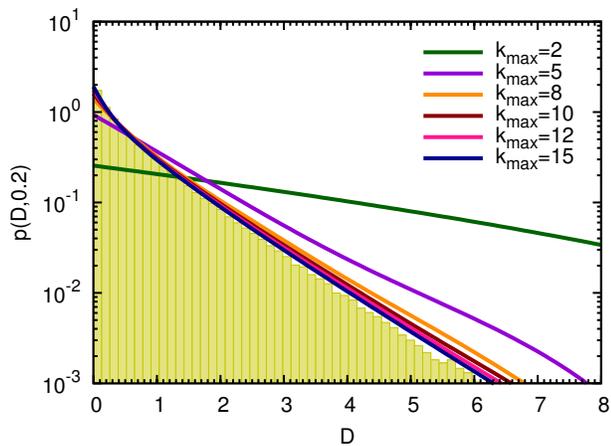}
\caption{\label{fig:pdssd vs inverse Fourier with bounded k window full}Same situation as in \Figref{fig:pdssd vs inverse Fourier with bounded k}, but the densities are now obtained by replacing the sharp cut-off at $k=\kmax$ by a smooth cut-off resulting from applying a half Hann window. A considerable improvement is achieved, especially if the value $\kmax$ is not too small.}
\end{figure}

Since the oscillations in the approximate densities of \Figref{fig:pdssd vs inverse Fourier with bounded k} seem to be induced by the hard cut-off at the wavelength $k=\kmax$, a possible strategy in reducing these oscillations may lie in applying an appropriate window function as in spectrum estimation procedures. We tested this option by applying a half Hann window to smoothen the cut-off. The best results were obtained for a window decaying from the value one at $k=0$ to zero at $k=\kmax$. The obtained results are very convincing if the cut-off value $\kmax$ is not too small as can be seen in \Figref{fig:pdssd vs inverse Fourier with bounded k window full}.

\section{\label{sec:Conclusions}Conclusions}

We investigated the connection between the signal attenuation measured by pulsed field gradient nuclear magnetic resonance and the distribution of single-particle diffusivities obtained from single-particle tracking. Due to their interrelations with the diffusion propagator of the system, the distribution of diffusivities is expressed by a general transformation of the signal attenuation. In the special case of a system involving two different states of diffusive mobility, the two-region exchange model of PFG NMR offers analytical expressions and allows for a comparison of analytical and simulated data. An approximation of the distribution of single-particle diffusivities via two populations with relative weights avoids the inverse Fourier transformation. Even in this simple system, such an approximation will only yield appropriate results if the time lag is much larger or much smaller than the dwell times. These cases correspond to an observation of the mean diffusion of the system and a process of completely separated diffusive motion without transition between the regions, respectively. Thus, in general, to obtain a proper distribution of single-particle diffusivities for diffusion in two-region systems, the exact transformation of the respective NMR signal attenuations is necessary. Only in this way we found perfect agreement of the experimental and analytical data. However, since PFG NMR data in some systems cannot be measured over a sufficiently large dynamic range, the inverse Fourier transformation may introduce deviations and oscillations. In these cases, the data analysis has to be performed with care and may require the use of additional information.

In summary, the investigated connection between two popular methods to experimentally observe and analyze diffusive motion offers new approaches for the evaluation of data. Hence, the methods of analysis may benefit from each other. This becomes especially relevant for systems with heterogeneities, where the distribution of diffusivities exhibits a dependence on the time lag. For more elaborated processes it may even not become stationary and enables to assess non-trivial properties of such systems. Since the distribution of diffusivities can be measured easily and contains more information from the propagator than well-established methods it should be used for future analysis of experimental data.

\begin{acknowledgments}
We gratefully acknowledge financial support from the Deutsche Forschungsgemeinschaft (DFG) for funding of the research unit FOR 877 ``From Local Constraints to Macroscopic Transport''. We also thank the anonymous referees for their valuable suggestions, which helped to improve the paper considerably.
\end{acknowledgments}

\appendix

\section{\label{sec:Appendix Correspondence iisf and signal attenuation}Correspondence between incoherent intermediate scattering function and signal attenuation}

The signal attenuation of PFG NMR and the incoherent intermediate scattering function as well as the dynamic structure factor are closely related. In this appendix, their correspondence is illustrated briefly and further details can be found in Refs.~\onlinecite{boon1991,hansen2006,kaerger1988}.

The observed motion of tracer particles can be analyzed by the self part of the van Hove time-dependent pair correlation function
\begin{equation}
 \label{eqn:definition van Hove correlation function}
\GselfVektor = \left\langle \frac{1}{N} \sum\limits_{i=1}^{N} \delta\Bigl( \disv-\bigl(\posv_{i}(\tau)-\posv_{i}(0)\bigr) \Bigr) \right\rangle
\end{equation}
describing the correlation of $N$ individual particles \cite{hove1954}. Its spatial Fourier transformation
\begin{equation}
 \label{eqn:Fourier transform of van Hove function is intermediate scattering function}
\iisf = \int \intdbegin{\disv} \GselfVektor \exp(\imag \vektor{k} \disv)
\end{equation}
leads to the incoherent intermediate scattering function
\begin{equation}
 \label{eqn:incoherent intermediate scattering function}
 \iisf = \frac{1}{N} \sum\limits_{i=1}^{N} \bigl\langle \exp(\imag \vektor{k} (\posv_{i}(\tau)-\posv_{i}(0))) \bigr\rangle
\text{,}
\end{equation}
which is linked to the velocity autocorrelation function of the particles. Furthermore, the incoherent intermediate scattering function $\iisf$ is related to the dynamic structure factor $S(\vektor{k},\omega)$ known from neutron scattering via Fourier transformation in $\tau$, i.e., the power spectrum of $\iisf$, where $\omega$ denotes a frequency.

For ergodic systems, $S(\vektor{k},\tau)$ can be obtained from an arbitrary particle
\begin{eqnarray}
\iisf & = & \bigl\langle \exp(\imag \vektor{k} (\posv(\tau)-\posv(0))) \bigr\rangle \nonumber \\
 \label{eqn:incoherent intermediate scattering function arbitrary particle with integral}
& = & \frac{1}{V^2} \iint\limits \intdbegin{\posv}\intd{\varPrime{\posv}} \exp(\imag \vektor{k} (\posv-\varPrime{\posv})\, p(\posv,\tau,\varPrime{\posv},0)
\qquad
\end{eqnarray}
where $V$ is the normalization and $p(\posv,\tau,\varPrime{\posv},0)$ denotes the joint probability of a particle to be located initially at $\varPrime{\posv}$ and at time $\tau$ at position $\posv$. The joint probability can be expressed by the conditional probability
\begin{equation}
 \label{eqn:joint probability vs conditional probability}
p(\posv,\tau,\varPrime{\posv},0) = p(\posv,\tau \vert \varPrime{\posv},0)\,p_{0}(\varPrime{\posv})
\text{.}
\end{equation}
Since during time $\tau$ the particle accomplished a displacement $\disv$, its positions are interrelated by $\posv=\varPrime{\posv}+\disv$. Due to translation invariance, without loss of generality, $\varPrime{\posv}=\vektor{0}$ leads to the propagator in Fourier space
\begin{equation}
 \label{eqn:propagator in Fourier space from iisf}
\frac{1}{V} \int\limits \intdbegin{\disv} \exp(\imag \vektor{k} \disv)\, p(\disv,\tau) = \PsiVektor(\tau,\vektor{k})
\end{equation}
corresponding to the signal attenuation in PFG NMR as introduced in \Eqref{eqn:signal attenuation Fourier transform of mean propagator}. Hence, signal attenuation and incoherent intermediate scattering function coincide. Furthermore, for identical particles without restrictions by the boundaries the averaging over the particles in \Eqref{eqn:definition van Hove correlation function} can be omitted and $\GselfVektor$ is equal to $p(\disv,\tau)$ given by \Eqref{eqn:mean propagator from ensemble average}.

For isotropic systems the self part of the radial van Hove time-dependent pair correlation function
\begin{equation}
\label{eqn:definition isotropic van Hove correlation function}
\GselfSkalar = \left\langle \frac{1}{N} \sum\limits_{i=1}^{N} \delta\left( \dis-|\posv_{i}(\tau)-\posv_{i}(0)| \right) \right\rangle
\end{equation}
considers only absolute values of the displacements. Again, for identical particles without restrictions by the boundaries an arbitrary particle can be considered and $\GselfSkalar$ is equal to $\pradial(\dis,\tau)$.

\section{\label{sec:Appendix Relation between evolution equations and signal attenuation}Relation between evolution equations and PFG NMR signal attenuation}

For \Eqref{eqn:equations of motion}, i.e., the evolution equations of the probability density to find a particle at position $\posv$ at time $t$, the moments of the random variable $\posv$ can be obtained via the characteristic functions. By introducing the vector $\vektor{p}(\vektor{k},t)$ comprising the characteristic functions of each region and the matrix $\mat{W}(\vektor{k})$ consisting of the elements
\begin{equation}
 \label{eqn:elements of matrix W}
\mat{W}(\vektor{k})_{nm} = w_{nm} + \left(-D_n \vektor{k}^2 - \sum\limits_l w_{ln}\right) \delta_{nm}
\text{,}
\end{equation}
the Fourier transform of \Eqref{eqn:equations of motion} can be written elegantly as
\begin{equation}
 \label{eqn:simplified Fourier transformed equation of motion}
 \frac{\intd}{\intd t} \vektor{p}(\vektor{k},t)= \mat{W}(\vektor{k})\, \vektor{p}(\vektor{k},t)
\end{equation}
where
\begin{equation}
 \label{eqn:solution of simplified Fourier transformed equation of motion}
\vektor{p}(\vektor{k},t) = \exp(t\, \mat{W}(\vektor{k}))\, \vektor{p}(\vektor{k},0)
\end{equation}
is easily seen to be the solution. For the two-region system the initial distribution $\vektor{p}(\vektor{k},0) = (\pi_1,\pi_2)\transpose$ is given by the equilibrium distribution between the regions.

Applying the spectral decomposition the matrix exponential in \Eqref{eqn:solution of simplified Fourier transformed equation of motion} for the two-region system can be written as
\begin{equation}
 \label{eqn:spectral decomposition of matrix exponential}
\exp(t\, \mat{W}(\vektor{k})) = \sum\limits_{\alpha=1}^2 \exp(t \mu_\alpha(\vektor{k}))\, \mat{A}_\alpha(\vektor{k})
\text{,}
\end{equation}
where
\begin{equation}
 \label{eqn:eigenvalues of spectral decomposition of matrix exponential}
\mu_{1,2}(\vektor{k}) = \frac{1}{2}(-\Dik{1} - \Dik{2} - \lambda \pm D(\vektor{k}))
\end{equation}
denote the eigenvalues and
\begin{equation}
 \label{eqn:matrices from eigenvectors of spectral decomposition of matrix exponential}
\mat{A}_{1,2}(\vektor{k}) = \frac{1}{2 D(\vektor{k})} \begin{pmatrix}
D(\vektor{k}) \pm \eta(\vektor{k})	& \pm 2w_{12} \\
\pm 2w_{21}	& D(\vektor{k}) \mp \eta(\vektor{k})
\end{pmatrix}
\end{equation}
represent the corresponding matrices from the dyadic product of the right- and left-eigenvectors with
\begin{eqnarray}
 \label{eqn:lambda in eigenvalues of spectral decomposition of matrix exponential}
\lambda & = & w_{21} + w_{12} \text{, } \\
 \label{eqn:eta in matrices of spectral decomposition of matrix exponential}
\eta(\vektor{k}) & = & -\Dik{1} + \Dik{2} - w_{21}+w_{12} \text{, and} \\
 \label{eqn:D in spectral decomposition of matrix exponential}
D(\vektor{k}) & = & \{(\Dik{1} + \Dik{2} + \lambda)^2 - 4 D_1 D_2 \vektor{k}^4 \nonumber \\*
		 && - 4 \Dik{1} w_{12} - 4 \Dik{2} w_{21}\}^{\frac{1}{2}}
\text{.}
\end{eqnarray}

Finally, the signal attenuation obtained from PFG NMR corresponds to the projection of the characteristic function
\begin{equation}
 \label{eqn:signal attenuation from projection of characteristic function}
\PsiVektor(\tau,\vektor{k}) = \begin{pmatrix} 1 & 1 \end{pmatrix} \exp(\tau\,\mat{W}(\vektor{k})) \begin{pmatrix} \pi_1 \\ \pi_2 \end{pmatrix}
\text{,}
\end{equation}
where $\vektor{k} = \kvarbitrary$ is measured in the direction of the applied field gradient denoted by the unit vector $\kUnitVektor$. Since for isotropic systems an arbitrary direction can be considered, \Eqref{eqn:signal attenuation from projection of characteristic function} results in the expressions given in \EqsToref{eqn:two-region system signal attenuation}{eqn:primed probability} for the two-region system.

\section{\label{sec:Appendix Exact transformation of limiting cases}Exact transformation of limiting cases}

By choosing $\vektor{k}=\kvarbitrary$, the isotropic signal attenuations for dimensionality $d$ in \EqsToref{eqn:transformation 1d}{eqn:transformation 3d}
\begin{equation*}
\PsiIsotropic\left(\tau,\frac{k}{u\sqrt{\tau}}\right) \text{ with }
u=\begin{cases}
    \sqrt{2} & \text{for } d=1 \\
	2 & \text{for } d=2 \\
	\sqrt{6} & \text{for } d=3
   \end{cases}
\text{,}
\end{equation*}
are considered in an arbitrary direction of the applied field gradient with intensity $k$. The exponent of \Eqref{eqn:signal attenuation from projection of characteristic function} is given by
\begin{equation}
 \label{eqn:exponent in projection of characteristic function}
\tau\,\mat{W}\left(\frac{\kvarbitrary}{u\sqrt{\tau}}\right) =%
\tau \begin{pmatrix}
 -w_{21} & w_{12} \\
 w_{21} & -w_{12}
\end{pmatrix}%
- \frac{k^2}{u^2} \begin{pmatrix}
   D_{1} & 0 \\
   0 & D_{2}
  \end{pmatrix}
\text{.}
\end{equation}
Based on these expressions the limiting cases are discussed separately.

\subsection*{Limiting case \texorpdfstring{$\tau \to 0$}{tau to 0}}
In the limiting case of $\tau \to 0$, only the diagonal matrix on the right hand side of \Eqref{eqn:exponent in projection of characteristic function} remains. Hence, the matrix exponential can be expressed by the exponentiation of the diagonal elements and \Eqref{eqn:signal attenuation from projection of characteristic function} reduces to
\begin{eqnarray}
&& \PsiIsotropic\left(\tau,\frac{k}{u\sqrt{\tau}}\right) \nonumber \\*
 \label{eqn:signal attenuation from projection of characteristic function in limiting case tau equal 0}
&& = \begin{pmatrix} 1 & 1 \end{pmatrix} \begin{pmatrix} \exp\left(-\frac{k^2}{u^2} D_1\right) & 0 \\ 0 & \exp\left(-\frac{k^2}{u^2} D_2\right) \end{pmatrix} \begin{pmatrix} \pi_1 \\ \pi_2 \end{pmatrix}
\qquad
\end{eqnarray}
yielding a superposition of two exponentials corresponding to separated regions. This is in agreement with previous findings since for short times $\tau$ no exchange between the regions occurs. Obviously, this result is not restricted to the two-region system but holds for an arbitrary number of diffusion states.

Applying the presented transformations \EqsToref{eqn:transformation 1d}{eqn:transformation 3d} for dimensionality $d$ to the obtained signal attenuation results in a distribution of diffusivities which is the superposition of two distributions of diffusivities for homogeneous diffusion in each region as given by \Eqref{eqn:homogeneous probability density of single-particle diffusivities general}, respectively.

\subsection*{Limiting case \texorpdfstring{$\tau \to \infty$}{tau to infinity}}
In the limiting case of $\tau \to \infty$, the situation is more complicated. Arguing analogously to the case of $\tau \to 0$ does not result in an appropriate expression. If the diagonal matrix on the right hand side of \Eqref{eqn:exponent in projection of characteristic function} is neglected, the signal attenuation will reduce to $1$ yielding only its normalization. Hence, this limiting case is addressed by involving the spectral decomposition. The matrices \Eqref{eqn:matrices from eigenvectors of spectral decomposition of matrix exponential} are given by
\begin{eqnarray}
&& \mat{A}_{1,2}\left(\frac{\kvarbitrary}{u\sqrt{\tau}}\right) \xrightarrow{\tau \to \infty} \mat{A}_{1,2}(\vektor{0}) \nonumber \\
 \label{eqn:spectral decomposition matrix A1 for 0}
&& \mat{A}_{1}(\vektor{0}) = \frac{1}{\lambda} \begin{pmatrix} w_{12} & w_{12} \\ w_{21} & w_{21} \end{pmatrix} \\
 \label{eqn:spectral decomposition matrix A2 for 0}
&& \mat{A}_{2}(\vektor{0}) = \frac{1}{\lambda} \begin{pmatrix} w_{21} & -w_{12} \\ -w_{21} & w_{12} \end{pmatrix}
\text{.}
\end{eqnarray}
Due to the projection in the signal attenuation \Eqref{eqn:signal attenuation from projection of characteristic function}
\begin{equation}
 \label{eqn:vanishing contribution from A2 for 0 due to projection}
\begin{pmatrix} 1 & 1 \end{pmatrix} \mat{A}_{2}(\vektor{0}) = \begin{pmatrix} 0 & 0 \end{pmatrix}
\end{equation}
the contribution from $\mat{A}_{2}(\vektor{0})$ vanishes. Thus, for $\tau \to \infty$ only eigenvalue $\mu_{1}$ contributes to the spectral decomposition. Moreover, $\mu_{1}=0$, which explains that the contribution from the diagonal matrix in \Eqref{eqn:exponent in projection of characteristic function} cannot be neglected.

Then, according to \Eqref{eqn:spectral decomposition of matrix exponential}, the exponential of $\tau\, \mu_{1}\left(\kvarbitrary/(u\sqrt{\tau})\right)$ is required, which is given by
\begin{equation}
 \label{eqn:tau times mu1 parameterized version}
\tau\, \mu_{1}\left(\frac{\kvarbitrary}{u\sqrt{\tau}}\right) = \frac{1}{2} \left(-a -\lambda \tau + \sqrt{(a + \lambda \tau)^{2} - b - c \tau} \right)
\end{equation}
with
\begin{subequations}
 \label{eqn:parameters for equation tau times mu1}
\begin{eqnarray}
\label{eqn:parameter a for equation tau times mu1}
a & = & D_{1} \frac{k^2}{u^2} + D_{2} \frac{k^2}{u^2} \\
\label{eqn:parameter b for equation tau times mu1}
b & = & 4 D_{1} D_{2} \frac{k^4}{u^4} \\
\label{eqn:parameter c for equation tau times mu1}
c & = & \left(4 D_{1} w_{12} + 4 D_{2} w_{21} \right) \frac{k^2}{u^2}
\text{.}
\end{eqnarray}
\end{subequations}
The square root in \Eqref{eqn:tau times mu1 parameterized version} can be rewritten as
\begin{eqnarray}
 && \sqrt{(a + \lambda \tau)^{2} - b - c \tau} \nonumber \\*
& = & \lambda \tau \sqrt{1 + \left(\frac{2a}{\lambda} - \frac{c}{\lambda^2}\right) \frac{1}{\tau} + \frac{a^2-b}{\lambda^2} \frac{1}{\tau^2}} \nonumber \\*
& = & \lambda \tau \left(1 + \frac{1}{2}\left(\frac{2a}{\lambda} - \frac{c}{\lambda^2}\right) \frac{1}{\tau} + \bigO{\frac{1}{\tau^2}} \right)
\label{eqn:Taylor expansion of square root in tau times mu1 parameterized version}
\text{.}
\end{eqnarray}
After further simplification, \Eqref{eqn:tau times mu1 parameterized version} reduces to
\begin{eqnarray}
\tau\, \mu_{1}\left(\frac{\kvarbitrary}{u\sqrt{\tau}}\right) & \simeq & \frac{1}{2} \left(-a -\lambda \tau + \lambda \tau + a - \frac{c}{2\lambda} \right) \nonumber \\*
 \label{eqn:tau times mu1 approximation with parameter}
& = & - \frac{c}{4\lambda}
\text{,}
\end{eqnarray}
which results in 
\begin{eqnarray}
\tau\, \mu_{1}\left(\frac{\kvarbitrary}{u\sqrt{\tau}}\right) & \simeq & -(\pi_1 D_1 + \pi_2 D_2) \frac{k^2}{u^2} \nonumber \\*
 \label{eqn:tau times mu1 approximation with mean diffusion coefficient}
& = & - \meanD \frac{k^2}{u^2}
\end{eqnarray}
by applying \Eqref{eqn:parameter c for equation tau times mu1} and \EqsAndref{eqn:stationary distribution between regions}{eqn:two-region system mean diffusivity}. Hence in the limiting case of $\tau \to \infty$, the signal attenuation
\begin{equation}
 \label{eqn:signal attenuation in limiting case tau to infinity}
\PsiIsotropic\left(\tau,\frac{k}{u\sqrt{\tau}}\right) = \exp\left(-\meanD \frac{k^2}{u^2}\right)
\end{equation}
depends only on the mean diffusion coefficient of the two-region system.

By integrating the signal attenuation \Eqref{eqn:signal attenuation in limiting case tau to infinity} for the limiting case $\tau \to \infty$ with the presented transformations \EqsToref{eqn:transformation 1d}{eqn:transformation 3d} for dimensionality $d$, as expected, the respective distributions of diffusivities \Eqref{eqn:homogeneous probability density of single-particle diffusivities general} are obtained, which correspond to homogeneous diffusion with the mean diffusion coefficient $\meanD$.

To conclude, the derivation of the two limiting cases reveals the properties of the distribution of single-particle diffusivities and its dependence on $\tau$. Starting from the limiting case $\tau \to \infty$, where only eigenvalue $\mu_{1}$ contributes, the weight of $\mu_{2}$ increases for decreasing $\tau$. This is reflected in the distribution of diffusivities by the dependence on $\tau$ as presented in \Figref{fig:pdssd vs complete inverse Fourier}. It describes the transition from a mean diffusion process to two completely separated diffusion processes for $\tau \to \infty$ and $\tau \to 0$, respectively. It should be noted that for the self part of the van Hove function the limiting cases cannot be determined. However, for the distribution of diffusivities, which is a rescaled van Hove self-correlation function, both limits are well-defined.

\raggedright
\bibliography{references}

\begin{thebibliography}{39}%
\makeatletter
\providecommand \@ifxundefined [1]{%
 \@ifx{#1\undefined}
}%
\providecommand \@ifnum [1]{%
 \ifnum #1\expandafter \@firstoftwo
 \else \expandafter \@secondoftwo
 \fi
}%
\providecommand \@ifx [1]{%
 \ifx #1\expandafter \@firstoftwo
 \else \expandafter \@secondoftwo
 \fi
}%
\providecommand \natexlab [1]{#1}%
\providecommand \enquote  [1]{``#1''}%
\providecommand \bibnamefont  [1]{#1}%
\providecommand \bibfnamefont [1]{#1}%
\providecommand \citenamefont [1]{#1}%
\providecommand \href@noop [0]{\@secondoftwo}%
\providecommand \href [0]{\begingroup \@sanitize@url \@href}%
\providecommand \@href[1]{\@@startlink{#1}\@@href}%
\providecommand \@@href[1]{\endgroup#1\@@endlink}%
\providecommand \@sanitize@url [0]{\catcode `\\12\catcode `\$12\catcode
  `\&12\catcode `\#12\catcode `\^12\catcode `\_12\catcode `\%12\relax}%
\providecommand \@@startlink[1]{}%
\providecommand \@@endlink[0]{}%
\providecommand \url  [0]{\begingroup\@sanitize@url \@url }%
\providecommand \@url [1]{\endgroup\@href {#1}{\urlprefix }}%
\providecommand \urlprefix  [0]{URL }%
\providecommand \Eprint [0]{\href }%
\providecommand \doibase [0]{http://dx.doi.org/}%
\providecommand \selectlanguage [0]{\@gobble}%
\providecommand \bibinfo  [0]{\@secondoftwo}%
\providecommand \bibfield  [0]{\@secondoftwo}%
\providecommand \translation [1]{[#1]}%
\providecommand \BibitemOpen [0]{}%
\providecommand \bibitemStop [0]{}%
\providecommand \bibitemNoStop [0]{.\EOS\space}%
\providecommand \EOS [0]{\spacefactor3000\relax}%
\providecommand \BibitemShut  [1]{\csname bibitem#1\endcsname}%
\let\auto@bib@innerbib\@empty
\bibitem [{\citenamefont {Heitjans}\ and\ \citenamefont
  {Kärger}(2005)}]{heitjans2005}%
  \BibitemOpen
  \bibinfo {editor} {\bibfnamefont {P.}~\bibnamefont {Heitjans}}\ and\ \bibinfo
  {editor} {\bibfnamefont {J.}~\bibnamefont {Kärger}},\ eds.,\ \href {\doibase
  10.1007/3-540-30970-5} {\emph {\bibinfo {title} {Diffusion in Condensed
  Matter}}},\ \bibinfo {edition} {2nd}\ ed.\ (\bibinfo  {publisher}
  {Springer},\ \bibinfo {address} {Berlin},\ \bibinfo {year}
  {2005})\BibitemShut {NoStop}%
\bibitem [{\citenamefont {Price}(2009)}]{price2009}%
  \BibitemOpen
  \bibfield  {author} {\bibinfo {author} {\bibfnamefont {W.~S.}\ \bibnamefont
  {Price}},\ }\href {\doibase 10.1017/CBO9780511770487} {\emph {\bibinfo
  {title} {NMR Studies of Translational Motion}}},\ Cambridge Molecular
  Science\ (\bibinfo  {publisher} {Cambridge University Press},\ \bibinfo
  {address} {Cambridge, New York},\ \bibinfo {year} {2009})\BibitemShut
  {NoStop}%
\bibitem [{\citenamefont {Krutyeva}\ and\ \citenamefont
  {Kärger}(2008)}]{krutyeva2008}%
  \BibitemOpen
  \bibfield  {author} {\bibinfo {author} {\bibfnamefont {M.}~\bibnamefont
  {Krutyeva}}\ and\ \bibinfo {author} {\bibfnamefont {J.}~\bibnamefont
  {Kärger}},\ }\href {\doibase 10.1021/la801426f} {\bibfield  {journal}
  {\bibinfo  {journal} {Langmuir}\ }\textbf {\bibinfo {volume} {24}},\ \bibinfo
  {pages} {10474} (\bibinfo {year} {2008})}\BibitemShut {NoStop}%
\bibitem [{\citenamefont {Nilsson}\ \emph {et~al.}(2010)\citenamefont
  {Nilsson}, \citenamefont {Alerstam}, \citenamefont {Wirestam}, \citenamefont
  {Staohlberg}, \citenamefont {Brockstedt},\ and\ \citenamefont
  {Lätt}}]{nilsson2010}%
  \BibitemOpen
  \bibfield  {author} {\bibinfo {author} {\bibfnamefont {M.}~\bibnamefont
  {Nilsson}}, \bibinfo {author} {\bibfnamefont {E.}~\bibnamefont {Alerstam}},
  \bibinfo {author} {\bibfnamefont {R.}~\bibnamefont {Wirestam}}, \bibinfo
  {author} {\bibfnamefont {F.}~\bibnamefont {Staohlberg}}, \bibinfo {author}
  {\bibfnamefont {S.}~\bibnamefont {Brockstedt}}, \ and\ \bibinfo {author}
  {\bibfnamefont {J.}~\bibnamefont {Lätt}},\ }\href {\doibase
  10.1016/j.jmr.2010.06.002} {\bibfield  {journal} {\bibinfo  {journal} {J.
  Magn. Reson.}\ }\textbf {\bibinfo {volume} {206}},\ \bibinfo {pages} {59}
  (\bibinfo {year} {2010})}\BibitemShut {NoStop}%
\bibitem [{\citenamefont {Saxton}\ and\ \citenamefont
  {Jacobson}(1997)}]{Saxton1997}%
  \BibitemOpen
  \bibfield  {author} {\bibinfo {author} {\bibfnamefont {M.~J.}\ \bibnamefont
  {Saxton}}\ and\ \bibinfo {author} {\bibfnamefont {K.}~\bibnamefont
  {Jacobson}},\ }\href {\doibase 10.1146/annurev.biophys.26.1.373} {\bibfield
  {journal} {\bibinfo  {journal} {Annu. Rev. Biophys. Biomol. Struct.}\
  }\textbf {\bibinfo {volume} {26}},\ \bibinfo {pages} {373} (\bibinfo {year}
  {1997})}\BibitemShut {NoStop}%
\bibitem [{\citenamefont {Zürner}\ \emph {et~al.}(2007)\citenamefont
  {Zürner}, \citenamefont {Kirstein}, \citenamefont {Döblinger},
  \citenamefont {Bräuchle},\ and\ \citenamefont {Bein}}]{Zurner2007}%
  \BibitemOpen
  \bibfield  {author} {\bibinfo {author} {\bibfnamefont {A.}~\bibnamefont
  {Zürner}}, \bibinfo {author} {\bibfnamefont {J.}~\bibnamefont {Kirstein}},
  \bibinfo {author} {\bibfnamefont {M.}~\bibnamefont {Döblinger}}, \bibinfo
  {author} {\bibfnamefont {C.}~\bibnamefont {Bräuchle}}, \ and\ \bibinfo
  {author} {\bibfnamefont {T.}~\bibnamefont {Bein}},\ }\href {\doibase
  10.1038/nature06398} {\bibfield  {journal} {\bibinfo  {journal} {Nature}\
  }\textbf {\bibinfo {volume} {450}},\ \bibinfo {pages} {705} (\bibinfo {year}
  {2007})}\BibitemShut {NoStop}%
\bibitem [{\citenamefont {Yu}\ \emph {et~al.}(1999)\citenamefont {Yu},
  \citenamefont {Richter}, \citenamefont {Datta}, \citenamefont {Durbin},\ and\
  \citenamefont {Dutta}}]{yu1999}%
  \BibitemOpen
  \bibfield  {author} {\bibinfo {author} {\bibfnamefont {C.-J.}\ \bibnamefont
  {Yu}}, \bibinfo {author} {\bibfnamefont {A.~G.}\ \bibnamefont {Richter}},
  \bibinfo {author} {\bibfnamefont {A.}~\bibnamefont {Datta}}, \bibinfo
  {author} {\bibfnamefont {M.~K.}\ \bibnamefont {Durbin}}, \ and\ \bibinfo
  {author} {\bibfnamefont {P.}~\bibnamefont {Dutta}},\ }\href {\doibase
  10.1103/PhysRevLett.82.2326} {\bibfield  {journal} {\bibinfo  {journal}
  {Phys. Rev. Lett.}\ }\textbf {\bibinfo {volume} {82}},\ \bibinfo {pages}
  {2326} (\bibinfo {year} {1999})}\BibitemShut {NoStop}%
\bibitem [{\citenamefont {Dembo}\ and\ \citenamefont
  {Zeitouni}(1986)}]{dembo1986}%
  \BibitemOpen
  \bibfield  {author} {\bibinfo {author} {\bibfnamefont {A.}~\bibnamefont
  {Dembo}}\ and\ \bibinfo {author} {\bibfnamefont {O.}~\bibnamefont
  {Zeitouni}},\ }\href {\doibase 10.1016/0304-4149(86)90018-9} {\bibfield
  {journal} {\bibinfo  {journal} {Stochastic Process. Appl.}\ }\textbf
  {\bibinfo {volume} {23}},\ \bibinfo {pages} {91} (\bibinfo {year}
  {1986})}\BibitemShut {NoStop}%
\bibitem [{\citenamefont {Campillo}\ and\ \citenamefont
  {Gland}(1989)}]{campillo1989}%
  \BibitemOpen
  \bibfield  {author} {\bibinfo {author} {\bibfnamefont {F.}~\bibnamefont
  {Campillo}}\ and\ \bibinfo {author} {\bibfnamefont {F.~L.}\ \bibnamefont
  {Gland}},\ }\href {\doibase 10.1016/0304-4149(89)90041-0} {\bibfield
  {journal} {\bibinfo  {journal} {Stochastic Process. Appl.}\ }\textbf
  {\bibinfo {volume} {33}},\ \bibinfo {pages} {245} (\bibinfo {year}
  {1989})}\BibitemShut {NoStop}%
\bibitem [{\citenamefont {Das}\ \emph {et~al.}(2009)\citenamefont {Das},
  \citenamefont {Cairo},\ and\ \citenamefont {Coombs}}]{das2009}%
  \BibitemOpen
  \bibfield  {author} {\bibinfo {author} {\bibfnamefont {R.}~\bibnamefont
  {Das}}, \bibinfo {author} {\bibfnamefont {C.~W.}\ \bibnamefont {Cairo}}, \
  and\ \bibinfo {author} {\bibfnamefont {D.}~\bibnamefont {Coombs}},\ }\href
  {\doibase 10.1371/journal.pcbi.1000556} {\bibfield  {journal} {\bibinfo
  {journal} {PLoS Comput. Biol.}\ }\textbf {\bibinfo {volume} {5}},\ \bibinfo
  {pages} {e1000556} (\bibinfo {year} {2009})}\BibitemShut {NoStop}%
\bibitem [{\citenamefont {Rabiner}(1989)}]{rabiner1989}%
  \BibitemOpen
  \bibfield  {author} {\bibinfo {author} {\bibfnamefont {L.~R.}\ \bibnamefont
  {Rabiner}},\ }\href {\doibase 10.1109/5.18626} {\bibfield  {journal}
  {\bibinfo  {journal} {Proceedings of the IEEE}\ }\textbf {\bibinfo {volume}
  {77}},\ \bibinfo {pages} {257} (\bibinfo {year} {1989})}\BibitemShut
  {NoStop}%
\bibitem [{\citenamefont {Elliott}\ \emph {et~al.}(1995)\citenamefont
  {Elliott}, \citenamefont {Aggoun},\ and\ \citenamefont
  {Moore}}]{elliott1995}%
  \BibitemOpen
  \bibfield  {author} {\bibinfo {author} {\bibfnamefont {R.~J.}\ \bibnamefont
  {Elliott}}, \bibinfo {author} {\bibfnamefont {L.}~\bibnamefont {Aggoun}}, \
  and\ \bibinfo {author} {\bibfnamefont {J.~B.}\ \bibnamefont {Moore}},\ }\href
  {\doibase 10.1007/978-0-387-84854-9} {\emph {\bibinfo {title} {Hidden Markov
  Models}}},\ \bibinfo {series} {Stochastic Modelling and Applied Probability},
  Vol.~\bibinfo {volume} {29}\ (\bibinfo  {publisher} {Springer},\ \bibinfo
  {address} {New York},\ \bibinfo {year} {1995})\BibitemShut {NoStop}%
\bibitem [{\citenamefont {Heidernätsch}\ \emph {et~al.}(2009)\citenamefont
  {Heidernätsch}, \citenamefont {Bauer}, \citenamefont {Täuber},
  \citenamefont {Radons},\ and\ \citenamefont {von
  Borczyskowski}}]{heidernaetsch2009}%
  \BibitemOpen
  \bibfield  {author} {\bibinfo {author} {\bibfnamefont {M.}~\bibnamefont
  {Heidernätsch}}, \bibinfo {author} {\bibfnamefont {M.}~\bibnamefont
  {Bauer}}, \bibinfo {author} {\bibfnamefont {D.}~\bibnamefont {Täuber}},
  \bibinfo {author} {\bibfnamefont {G.}~\bibnamefont {Radons}}, \ and\ \bibinfo
  {author} {\bibfnamefont {C.}~\bibnamefont {von Borczyskowski}},\ }\href
  {http://www.uni-leipzig.de/diffusion/journal/pdf/volume11/diff_fund_11%28200%
9%29111.pdf} {\bibfield  {journal} {\bibinfo  {journal} {Diffus. Fundam.}\
  }\textbf {\bibinfo {volume} {11}},\ \bibinfo {pages} {111} (\bibinfo {year}
  {2009})}\BibitemShut {NoStop}%
\bibitem [{\citenamefont {Bauer}\ \emph {et~al.}(2009)\citenamefont {Bauer},
  \citenamefont {Heidernätsch}, \citenamefont {Täuber}, \citenamefont {von
  Borczyskowski},\ and\ \citenamefont {Radons}}]{bauer2009}%
  \BibitemOpen
  \bibfield  {author} {\bibinfo {author} {\bibfnamefont {M.}~\bibnamefont
  {Bauer}}, \bibinfo {author} {\bibfnamefont {M.}~\bibnamefont
  {Heidernätsch}}, \bibinfo {author} {\bibfnamefont {D.}~\bibnamefont
  {Täuber}}, \bibinfo {author} {\bibfnamefont {C.}~\bibnamefont {von
  Borczyskowski}}, \ and\ \bibinfo {author} {\bibfnamefont {G.}~\bibnamefont
  {Radons}},\ }\href
  {http://www.uni-leipzig.de/diffusion/journal/pdf/volume11/diff_fund_11%28200%
9%29104.pdf} {\bibfield  {journal} {\bibinfo  {journal} {Diffus. Fundam.}\
  }\textbf {\bibinfo {volume} {11}},\ \bibinfo {pages} {104} (\bibinfo {year}
  {2009})}\BibitemShut {NoStop}%
\bibitem [{\citenamefont {Saxton}(1997)}]{Saxton1997b}%
  \BibitemOpen
  \bibfield  {author} {\bibinfo {author} {\bibfnamefont {M.~J.}\ \bibnamefont
  {Saxton}},\ }\href {\doibase 10.1016/S0006-3495(97)78820-9} {\bibfield
  {journal} {\bibinfo  {journal} {Biophys. J.}\ }\textbf {\bibinfo {volume}
  {72}},\ \bibinfo {pages} {1744} (\bibinfo {year} {1997})}\BibitemShut
  {NoStop}%
\bibitem [{\citenamefont {Trenkmann}\ \emph {et~al.}(2009)\citenamefont
  {Trenkmann}, \citenamefont {Täuber}, \citenamefont {Bauer}, \citenamefont
  {Schuster}, \citenamefont {Bok}, \citenamefont {Gangopadhyay},\ and\
  \citenamefont {von Borczyskowski}}]{trenkmann2009}%
  \BibitemOpen
  \bibfield  {author} {\bibinfo {author} {\bibfnamefont {I.}~\bibnamefont
  {Trenkmann}}, \bibinfo {author} {\bibfnamefont {D.}~\bibnamefont {Täuber}},
  \bibinfo {author} {\bibfnamefont {M.}~\bibnamefont {Bauer}}, \bibinfo
  {author} {\bibfnamefont {J.}~\bibnamefont {Schuster}}, \bibinfo {author}
  {\bibfnamefont {S.}~\bibnamefont {Bok}}, \bibinfo {author} {\bibfnamefont
  {S.}~\bibnamefont {Gangopadhyay}}, \ and\ \bibinfo {author} {\bibfnamefont
  {C.}~\bibnamefont {von Borczyskowski}},\ }\href
  {http://www.uni-leipzig.de/diffusion/journal/pdf/volume11/diff_fund_11%28200%
9%29108.pdf} {\bibfield  {journal} {\bibinfo  {journal} {Diffus. Fundam.}\
  }\textbf {\bibinfo {volume} {11}},\ \bibinfo {pages} {108} (\bibinfo {year}
  {2009})}\BibitemShut {NoStop}%
\bibitem [{\citenamefont {Lubelski}\ \emph {et~al.}(2008)\citenamefont
  {Lubelski}, \citenamefont {Sokolov},\ and\ \citenamefont
  {Klafter}}]{lubelski2008}%
  \BibitemOpen
  \bibfield  {author} {\bibinfo {author} {\bibfnamefont {A.}~\bibnamefont
  {Lubelski}}, \bibinfo {author} {\bibfnamefont {I.~M.}\ \bibnamefont
  {Sokolov}}, \ and\ \bibinfo {author} {\bibfnamefont {J.}~\bibnamefont
  {Klafter}},\ }\href {\doibase 10.1103/PhysRevLett.100.250602} {\bibfield
  {journal} {\bibinfo  {journal} {Phys. Rev. Lett.}\ }\textbf {\bibinfo
  {volume} {100}},\ \bibinfo {pages} {250602} (\bibinfo {year}
  {2008})}\BibitemShut {NoStop}%
\bibitem [{\citenamefont {Kärger}\ \emph {et~al.}(1988)\citenamefont
  {Kärger}, \citenamefont {Pfeifer},\ and\ \citenamefont
  {Heink}}]{kaerger1988}%
  \BibitemOpen
  \bibfield  {author} {\bibinfo {author} {\bibfnamefont {J.}~\bibnamefont
  {Kärger}}, \bibinfo {author} {\bibfnamefont {H.}~\bibnamefont {Pfeifer}}, \
  and\ \bibinfo {author} {\bibfnamefont {W.}~\bibnamefont {Heink}},\ }in\
  \href@noop {} {\emph {\bibinfo {booktitle} {Advances in Magnetic
  Resonance}}},\ Vol.~\bibinfo {volume} {12},\ \bibinfo {editor} {edited by\
  \bibinfo {editor} {\bibfnamefont {J.~S.}\ \bibnamefont {Waugh}}}\ (\bibinfo
  {publisher} {Academic Press},\ \bibinfo {address} {San Diego},\ \bibinfo
  {year} {1988})\ pp.\ \bibinfo {pages} {1--89}\BibitemShut {NoStop}%
\bibitem [{\citenamefont {Van~den Broeck}\ and\ \citenamefont
  {Mazo}(1984)}]{vandenbroeck1984}%
  \BibitemOpen
  \bibfield  {author} {\bibinfo {author} {\bibfnamefont {C.}~\bibnamefont
  {Van~den Broeck}}\ and\ \bibinfo {author} {\bibfnamefont {R.~M.}\
  \bibnamefont {Mazo}},\ }\href {\doibase 10.1063/1.448110} {\bibfield
  {journal} {\bibinfo  {journal} {J. Chem. Phys.}\ }\textbf {\bibinfo {volume}
  {81}},\ \bibinfo {pages} {3624} (\bibinfo {year} {1984})}\BibitemShut
  {NoStop}%
\bibitem [{\citenamefont {Blümich}(2005)}]{bluemich2005}%
  \BibitemOpen
  \bibfield  {author} {\bibinfo {author} {\bibfnamefont {B.}~\bibnamefont
  {Blümich}},\ }\href {\doibase 10.1007/b138660} {\emph {\bibinfo {title}
  {Essential NMR}}},\ Vol.~\bibinfo {volume} {1}\ (\bibinfo  {publisher}
  {Springer},\ \bibinfo {address} {Berlin},\ \bibinfo {year}
  {2005})\BibitemShut {NoStop}%
\bibitem [{\citenamefont {Kimmich}\ \emph {et~al.}(2008)\citenamefont
  {Kimmich}, \citenamefont {Fatkullin}, \citenamefont {Kehr},\ and\
  \citenamefont {Li}}]{kimmich2008}%
  \BibitemOpen
  \bibfield  {author} {\bibinfo {author} {\bibfnamefont {R.}~\bibnamefont
  {Kimmich}}, \bibinfo {author} {\bibfnamefont {N.}~\bibnamefont {Fatkullin}},
  \bibinfo {author} {\bibfnamefont {M.}~\bibnamefont {Kehr}}, \ and\ \bibinfo
  {author} {\bibfnamefont {Y.}~\bibnamefont {Li}},\ }in\ \href {\doibase
  10.1002/9783527622979.ch17} {\emph {\bibinfo {booktitle} {Anomalous
  Transport--Foundations and Applications}}},\ \bibinfo {editor} {edited by\
  \bibinfo {editor} {\bibfnamefont {R.}~\bibnamefont {Klages}}, \bibinfo
  {editor} {\bibfnamefont {G.}~\bibnamefont {Radons}}, \ and\ \bibinfo {editor}
  {\bibfnamefont {I.~M.}\ \bibnamefont {Sokolov}}}\ (\bibinfo  {publisher}
  {Wiley-VCH},\ \bibinfo {address} {Berlin},\ \bibinfo {year} {2008})\ Chap.\
  \bibinfo {chapter} {XVII}, pp.\ \bibinfo {pages} {485--518}\BibitemShut
  {NoStop}%
\bibitem [{\citenamefont {Valiullin}\ and\ \citenamefont
  {Kärger}(2008)}]{valiullin2008}%
  \BibitemOpen
  \bibfield  {author} {\bibinfo {author} {\bibfnamefont {R.}~\bibnamefont
  {Valiullin}}\ and\ \bibinfo {author} {\bibfnamefont {J.}~\bibnamefont
  {Kärger}},\ }in\ \href {\doibase 10.1002/9783527622979.ch18} {\emph
  {\bibinfo {booktitle} {Anomalous Transport--Foundations and Applications}}},\
  \bibinfo {editor} {edited by\ \bibinfo {editor} {\bibfnamefont
  {R.}~\bibnamefont {Klages}}, \bibinfo {editor} {\bibfnamefont
  {G.}~\bibnamefont {Radons}}, \ and\ \bibinfo {editor} {\bibfnamefont {I.~M.}\
  \bibnamefont {Sokolov}}}\ (\bibinfo  {publisher} {Wiley-VCH},\ \bibinfo
  {address} {Berlin},\ \bibinfo {year} {2008})\ Chap.\ \bibinfo {chapter}
  {XVIII}, pp.\ \bibinfo {pages} {519--544}\BibitemShut {NoStop}%
\bibitem [{\citenamefont {Kärger}\ and\ \citenamefont
  {Heink}(1983)}]{kaerger1983}%
  \BibitemOpen
  \bibfield  {author} {\bibinfo {author} {\bibfnamefont {J.}~\bibnamefont
  {Kärger}}\ and\ \bibinfo {author} {\bibfnamefont {W.}~\bibnamefont
  {Heink}},\ }\href {\doibase 10.1016/0022-2364(83)90094-X} {\bibfield
  {journal} {\bibinfo  {journal} {J. Magn. Reson.}\ }\textbf {\bibinfo {volume}
  {51}},\ \bibinfo {pages} {1} (\bibinfo {year} {1983})}\BibitemShut {NoStop}%
\bibitem [{\citenamefont {Gardiner}(2004)}]{gardiner2004}%
  \BibitemOpen
  \bibfield  {author} {\bibinfo {author} {\bibfnamefont {C.~W.}\ \bibnamefont
  {Gardiner}},\ }\href@noop {} {\emph {\bibinfo {title} {Handbook of Stochastic
  Methods}}},\ \bibinfo {edition} {3rd}\ ed.,\ \bibinfo {series} {Springer
  Series in Synergetics}, Vol.~\bibinfo {volume} {13}\ (\bibinfo  {publisher}
  {Springer},\ \bibinfo {address} {Berlin},\ \bibinfo {year}
  {2004})\BibitemShut {NoStop}%
\bibitem [{\citenamefont {Mandelbrot}\ and\ \citenamefont
  {Van~Ness}(1968)}]{mandelbrot1968}%
  \BibitemOpen
  \bibfield  {author} {\bibinfo {author} {\bibfnamefont {B.~B.}\ \bibnamefont
  {Mandelbrot}}\ and\ \bibinfo {author} {\bibfnamefont {J.~W.}\ \bibnamefont
  {Van~Ness}},\ }\href {http://www.jstor.org/stable/2027184} {\bibfield
  {journal} {\bibinfo  {journal} {SIAM Review}\ }\textbf {\bibinfo {volume}
  {10}},\ \bibinfo {pages} {422} (\bibinfo {year} {1968})}\BibitemShut
  {NoStop}%
\bibitem [{\citenamefont {Klages}\ \emph {et~al.}(2008)\citenamefont {Klages},
  \citenamefont {Radons},\ and\ \citenamefont {Sokolov}}]{klages2008}%
  \BibitemOpen
  \bibinfo {editor} {\bibfnamefont {R.}~\bibnamefont {Klages}}, \bibinfo
  {editor} {\bibfnamefont {G.}~\bibnamefont {Radons}}, \ and\ \bibinfo {editor}
  {\bibfnamefont {I.~M.}\ \bibnamefont {Sokolov}},\ eds.,\ \href {\doibase
  10.1002/9783527622979} {\emph {\bibinfo {title} {Anomalous
  Transport--Foundations and Applications}}}\ (\bibinfo  {publisher}
  {Wiley-VCH},\ \bibinfo {address} {Berlin},\ \bibinfo {year}
  {2008})\BibitemShut {NoStop}%
\bibitem [{\citenamefont {Kusumi}\ \emph {et~al.}(1993)\citenamefont {Kusumi},
  \citenamefont {Sako},\ and\ \citenamefont {Yamamoto}}]{kusumi1993}%
  \BibitemOpen
  \bibfield  {author} {\bibinfo {author} {\bibfnamefont {A.}~\bibnamefont
  {Kusumi}}, \bibinfo {author} {\bibfnamefont {Y.}~\bibnamefont {Sako}}, \ and\
  \bibinfo {author} {\bibfnamefont {M.}~\bibnamefont {Yamamoto}},\ }\href
  {\doibase 10.1016/S0006-3495(93)81253-0} {\bibfield  {journal} {\bibinfo
  {journal} {Biophys. J.}\ }\textbf {\bibinfo {volume} {65}},\ \bibinfo {pages}
  {2021} (\bibinfo {year} {1993})}\BibitemShut {NoStop}%
\bibitem [{\citenamefont {Schuster}\ \emph {et~al.}(2007)\citenamefont
  {Schuster}, \citenamefont {Brabandt},\ and\ \citenamefont {von
  Borczyskowski}}]{schuster2007}%
  \BibitemOpen
  \bibfield  {author} {\bibinfo {author} {\bibfnamefont {J.}~\bibnamefont
  {Schuster}}, \bibinfo {author} {\bibfnamefont {J.}~\bibnamefont {Brabandt}},
  \ and\ \bibinfo {author} {\bibfnamefont {C.}~\bibnamefont {von
  Borczyskowski}},\ }\href {\doibase 10.1016/j.jlumin.2007.02.028} {\bibfield
  {journal} {\bibinfo  {journal} {J. Lumin.}\ }\textbf {\bibinfo {volume}
  {127}},\ \bibinfo {pages} {224} (\bibinfo {year} {2007})}\BibitemShut
  {NoStop}%
\bibitem [{\citenamefont {Sbalzarini}\ and\ \citenamefont
  {Koumoutsakosa}(2005)}]{sbalzarini2005}%
  \BibitemOpen
  \bibfield  {author} {\bibinfo {author} {\bibfnamefont {I.~F.}\ \bibnamefont
  {Sbalzarini}}\ and\ \bibinfo {author} {\bibfnamefont {P.}~\bibnamefont
  {Koumoutsakosa}},\ }\href {\doibase 10.1016/j.jsb.2005.06.002} {\bibfield
  {journal} {\bibinfo  {journal} {J. Struct. Biol.}\ }\textbf {\bibinfo
  {volume} {151}},\ \bibinfo {pages} {182} (\bibinfo {year}
  {2005})}\BibitemShut {NoStop}%
\bibitem [{\citenamefont {Schob}\ and\ \citenamefont
  {Cichos}(2006)}]{Schob2006}%
  \BibitemOpen
  \bibfield  {author} {\bibinfo {author} {\bibfnamefont {A.}~\bibnamefont
  {Schob}}\ and\ \bibinfo {author} {\bibfnamefont {F.}~\bibnamefont {Cichos}},\
  }\href {\doibase 10.1021/jp055201+} {\bibfield  {journal} {\bibinfo
  {journal} {J. Phys. Chem. B}\ }\textbf {\bibinfo {volume} {110}},\ \bibinfo
  {pages} {4354} (\bibinfo {year} {2006})}\BibitemShut {NoStop}%
\bibitem [{\citenamefont {van Kampen}(1992)}]{kampen1992}%
  \BibitemOpen
  \bibfield  {author} {\bibinfo {author} {\bibfnamefont {N.~G.}\ \bibnamefont
  {van Kampen}},\ }\href
  {http://www.sciencedirect.com/science/book/9780444529657} {\emph {\bibinfo
  {title} {Stochastic processes in physics and chemistry}}},\ \bibinfo
  {edition} {2nd}\ ed.\ (\bibinfo  {publisher} {North-Holland},\ \bibinfo
  {address} {Amsterdam},\ \bibinfo {year} {1992})\BibitemShut {NoStop}%
\bibitem [{\citenamefont {Haus}\ and\ \citenamefont {Kehr}(1987)}]{haus1987}%
  \BibitemOpen
  \bibfield  {author} {\bibinfo {author} {\bibfnamefont {J.~W.}\ \bibnamefont
  {Haus}}\ and\ \bibinfo {author} {\bibfnamefont {K.~W.}\ \bibnamefont
  {Kehr}},\ }\href {\doibase 10.1016/0370-1573(87)90005-6} {\bibfield
  {journal} {\bibinfo  {journal} {Phys. Rep.}\ }\textbf {\bibinfo {volume}
  {150}},\ \bibinfo {pages} {263} (\bibinfo {year} {1987})}\BibitemShut
  {NoStop}%
\bibitem [{\citenamefont {Weiss}(1994)}]{weiss1994}%
  \BibitemOpen
  \bibfield  {author} {\bibinfo {author} {\bibfnamefont {G.~H.}\ \bibnamefont
  {Weiss}},\ }\href@noop {} {\emph {\bibinfo {title} {Aspects and Applications
  of the Random Walk}}}\ (\bibinfo  {publisher} {North-Holland},\ \bibinfo
  {address} {Amsterdam},\ \bibinfo {year} {1994})\BibitemShut {NoStop}%
\bibitem [{\citenamefont {Malchus}\ and\ \citenamefont
  {Weiss}(2010)}]{malchus2010}%
  \BibitemOpen
  \bibfield  {author} {\bibinfo {author} {\bibfnamefont {N.}~\bibnamefont
  {Malchus}}\ and\ \bibinfo {author} {\bibfnamefont {M.}~\bibnamefont
  {Weiss}},\ }\href {\doibase 10.1016/j.bpj.2010.06.020} {\bibfield  {journal}
  {\bibinfo  {journal} {Biophys. J.}\ }\textbf {\bibinfo {volume} {99}},\
  \bibinfo {pages} {1321} (\bibinfo {year} {2010})}\BibitemShut {NoStop}%
\bibitem [{\citenamefont {Vembu}(1961)}]{vembu1961}%
  \BibitemOpen
  \bibfield  {author} {\bibinfo {author} {\bibfnamefont {S.}~\bibnamefont
  {Vembu}},\ }\href {\doibase 10.1093/qmath/12.1.165} {\bibfield  {journal}
  {\bibinfo  {journal} {Quart. J. Math.}\ }\textbf {\bibinfo {volume} {12}},\
  \bibinfo {pages} {165} (\bibinfo {year} {1961})}\BibitemShut {NoStop}%
\bibitem [{\citenamefont {Fleischer}\ and\ \citenamefont
  {Fujara}(1994)}]{fleischer1994}%
  \BibitemOpen
  \bibfield  {author} {\bibinfo {author} {\bibfnamefont {G.}~\bibnamefont
  {Fleischer}}\ and\ \bibinfo {author} {\bibfnamefont {F.}~\bibnamefont
  {Fujara}},\ }in\ \href@noop {} {\emph {\bibinfo {booktitle} {NMR - Basic
  Principles and Progress}}},\ Vol.~\bibinfo {volume} {30},\ \bibinfo {editor}
  {edited by\ \bibinfo {editor} {\bibfnamefont {P.}~\bibnamefont {Diehl}},
  \bibinfo {editor} {\bibfnamefont {E.}~\bibnamefont {Fluck}}, \bibinfo
  {editor} {\bibfnamefont {H.}~\bibnamefont {Günther}}, \bibinfo {editor}
  {\bibfnamefont {R.}~\bibnamefont {Kosfeld}}, \bibinfo {editor} {\bibfnamefont
  {J.}~\bibnamefont {Seelig}}, \ and\ \bibinfo {editor} {\bibfnamefont
  {B.}~\bibnamefont {Blümich}}}\ (\bibinfo  {publisher} {Springer-Verlag},\
  \bibinfo {address} {Berlin},\ \bibinfo {year} {1994})\ Chap.~\bibinfo
  {chapter} {IV}, pp.\ \bibinfo {pages} {159--207}\BibitemShut {NoStop}%
\bibitem [{\citenamefont {Boon}\ and\ \citenamefont {Yip}(1991)}]{boon1991}%
  \BibitemOpen
  \bibfield  {author} {\bibinfo {author} {\bibfnamefont {J.~P.}\ \bibnamefont
  {Boon}}\ and\ \bibinfo {author} {\bibfnamefont {S.}~\bibnamefont {Yip}},\
  }\href@noop {} {\emph {\bibinfo {title} {Molecular Hydrodynamics}}}\
  (\bibinfo  {publisher} {Dover Publications},\ \bibinfo {address} {New York},\
  \bibinfo {year} {1991})\BibitemShut {NoStop}%
\bibitem [{\citenamefont {Hansen}\ and\ \citenamefont
  {McDonald}(2006)}]{hansen2006}%
  \BibitemOpen
  \bibfield  {author} {\bibinfo {author} {\bibfnamefont {J.-P.}\ \bibnamefont
  {Hansen}}\ and\ \bibinfo {author} {\bibfnamefont {I.~R.}\ \bibnamefont
  {McDonald}},\ }\href
  {http://www.sciencedirect.com/science/book/9780123705358} {\emph {\bibinfo
  {title} {Theory of Simple Liquids}}},\ \bibinfo {edition} {3rd}\ ed.\
  (\bibinfo  {publisher} {Academic Press},\ \bibinfo {address} {Amsterdam},\
  \bibinfo {year} {2006})\BibitemShut {NoStop}%
\bibitem [{\citenamefont {van Hove}(1954)}]{hove1954}%
  \BibitemOpen
  \bibfield  {author} {\bibinfo {author} {\bibfnamefont {L.}~\bibnamefont {van
  Hove}},\ }\href {\doibase 10.1103/PhysRev.95.249} {\bibfield  {journal}
  {\bibinfo  {journal} {Phys. Rev.}\ }\textbf {\bibinfo {volume} {95}},\
  \bibinfo {pages} {249} (\bibinfo {year} {1954})}\BibitemShut {NoStop}%
\end{thebibliography}%

\end{document}